\documentclass[twocolumn,showpacs,preprintnumbers,amsmath,amssymb,superscriptaddress]{revtex4-1}
\usepackage[maxfloats=256]{morefloats}
\usepackage{graphicx}% Include figure files
\usepackage{dcolumn}% Align table columns on the decimal point
\usepackage{bm}% bold math 
\usepackage{tabularx}
\usepackage{longtable}
\usepackage{booktabs} % For better table lines
\usepackage{ulem}
\newcolumntype{M}{>{\centering\arraybackslash}m{1.85cm}}
\usepackage[export]{adjustbox}
\usepackage{float}
\usepackage{color}   %May be necessary if you want to color links
\makeatletter
\newcommand{\colorcaption}[2][]{%
  \begingroup%
  \renewcommand{\@caption@fignum@sep}{ (Color online). }%
  \caption[#1]{#2}%
  \endgroup%
}
\makeatother

\usepackage{hyperref}
\hypersetup{
    colorlinks=true,
    linkcolor=blue,
    filecolor=red,      
    urlcolor=cyan,
    pdfpagemode=FullScreen,
}



\newcommand\T{\rule{0pt}{3ex}}       % Top strut
\newcommand\B{\rule[-1.5ex]{0pt}{0pt}} % Bottom strut

\newcommand{\orcid}[1]{\href{https://orcid.org/#1}{\hskip2pt\includegraphics[width=9pt]{orcid-ID.png}}}
\begin{document}

	\title{Shell-model study for allowed and forbidden $\beta^-$ decay properties in the mass region ``south" of $^{208}$Pb}
	\author{Shweta  Sharma}
	\address{Department of Physics, Indian Institute of Technology Roorkee, Roorkee 247667, India}
	\author{Praveen C. Srivastava}
	\email{Corresponding author: praveen.srivastava@ph.iitr.ac.in}
	\address{Department of Physics, Indian Institute of Technology Roorkee, Roorkee 247667, India}
	\author{Anil Kumar}
        \address{Center for Computational Sciences, University of Tsukuba, 1-1-1 Tennodai, Tsukuba, Ibaraki 305-8577, Japan}
	\author{Toshio Suzuki}
	\address{Department of Physics, College of Humanities and Sciences, Nihon University, Sakurajosui 3, Setagaya-ku,  Tokyo 156-8550, Japan}
	\address{NAT Research Center, NAT Corporation, 3129-45 Hibara Muramatsu, Tokai-mura, Naka-gun, Ibaraki 319-1112, Japan}
	\address{School of Physics, Beihang University 100083, People's Republic of China}
	\author{Cenxi Yuan}
	\address{Sino-French Institute of Nuclear Engineering and Technology, Sun Yat-Sen University, Zhuhai, \\Guangdong 519082, China}
        \author{ Noritaka Shimizu}
        \address{Center for Computational Sciences, University of Tsukuba, 1-1-1 Tennodai, Tsukuba, Ibaraki 305-8577, Japan}

	\date{\hfill \today}
	%%%%%%%%%%%%%%%%%%%%%%%%%%%%%%%%%%%%%%%%%
	%\bibliographystyle{prsty}
	\begin{abstract}
		
		The large-scale shell-model calculations have been performed for the neutron-rich nuclei in the south region of $^{208}$Pb in the nuclear chart. The $\beta$-decay properties, such as  the $\log ft$, average shape factor values,  half-lives, and partial decay rates are calculated for these neutron-rich nuclei using recent effective interaction 
 for the $^{208}$Pb region.  These calculations have been performed without truncation in a particular model space for nuclei $N\leq 126$; additionally, particle-hole excitations are included in the case of core-breaking nuclei ($Z\leq 82, N>126$). An extensive comparison with the experimental data has been made, and spin parities of several states have been proposed.

	\end{abstract}
	
	\maketitle
	
	\section{Introduction}
	In the early universe, light elements up to mass number $A=7$ were produced by big-bang nucleosynthesis. When stars were formed later, nucleosynthesis started with the fusion of elements in the hot, dense environment inside the stars \cite{burbidge1957}. This process continues up until iron and the nuclei heavier than iron are produced via various astrophysical processes like $r$ (rapid neutron capture) \cite{kajino2019}, $s$ (slow neutron capture) \cite{kappeler1989}, and $p$ (proton capture) \cite{arnould2003} process.  The nuclei near the stability line in the nuclear chart undergo $s$-process nucleosynthesis, and the neutron-rich nuclei away from the stability line and near the neutron drip line undergo $r$-process nucleosynthesis. The nuclei around doubly magic nucleus $^{208}$Pb warrant extensive investigation due to their astrophysical importance \cite{suzuki2012,borzov2000}. Concurrently, $\beta$-decay properties of exotic nuclei near the $r$-process path, such as $\beta$ decay half-lives and $\beta$-delayed neutron emission probabilities are crucial to determining the relative $A=195$ $r$-process peak abundances \cite{mumpower2016,surman2014,watanabe2015,morales2014} and the matter flow to heavier nuclei \cite{nieto2014}.
%In the Pb region there is competition between allowed and forbidden transitions.

Of particular interest is the south region of the $^{208}$Pb, due to the experimental difficulties there is still a lack of the experimental $\beta$- decay half-lives for the neutron-rich nuclei around the $N=126$ and $Z\leq 80$. However, Morales {\it et al.} \cite{morales2014} measured the $\beta$-decay half-lives of 19 nuclei across the $N=126$ shell closure. More recently in Ref. \cite{Folch}, they have also measured the $\beta$-decay half-lives of the 20 nuclei in the lead region. Moreover, several new experimental plans are ongoing to measure the $\beta$-decay half-lives around the $N=126$ \cite{watanabe2015,cern1,RIKEN1,Even}.  Due to a lack of experimental information on $\beta$-decay properties, the theoretical prediction of the $\beta$-decay properties from the different models helps them to understand the underlying nuclear structure properties of these nuclei located on or around $r$-process path.  On the theoretical side, several systematic studies about the $\beta$-decay properties have been investigated by using the different nuclear models in the vicinity of $^{208}$Pb \cite{Moller, Moller2, Engel, Fang, Borzov, Borzov2, Marketin, Mustonen, Ney,Niu, Robin2014}. Among them, the less number of shell model studies have been performed to investigate the $\beta$-decays properties in the lead region. In Refs. \cite{Langanke, Pinedo}, only Gamow-Teller transitions have been involved in the $\beta$-decay rate calculations through the shell model. However, the contribution of first-forbidden transitions in addition to Gamow-Teller transitions also plays an important role in the total half-lives near $Z=82$ \cite{Borzov}. The shell model calculations have also been extended hereafter by including the first-forbidden transitions \cite{suzuki2012,zhi2013}. The enhancement of first-forbidden decays is expected due to the ordering of the proton and neutron orbitals below $^{208}$Pb \cite{Morales_JPC}. More recently, a large scale shell model study of the $\beta^-$decay properties of $N = 125$ and $126$ isotones with $Z = 52-79$ have been preformed by considering both Gamow-Teller and first-forbidden transitions \cite{Anil},  it is observed that the contributions of first-forbidden transitions are diminished with a decrease in proton number from $Z = 82$.

 The forbidden transitions compete with the allowed transitions, making forbidden transitions the major decay mode in the heavier mass region. To study this competition, nuclei should have a small number of both negative and positive parity levels below the $Q$-value. In order to test this competition, Carroll \textit{et al.} \cite{car2020} studied the $\beta$ decay of $^{208}$Hg into the one-proton-hole and one-neutron-particle nucleus $^{208}$Tl. Further, Brunet \textit{et al.} \cite{brunet2021} studied $\beta$ decay of nuclei in the $N<126$, $Z>82$ region, i.e., $^{208}$At into $^{208}$Po transitions.  Thus, the $\beta$ decay study provides a major opportunity to address this quest.

Due to the interaction between the single particle and the collective degree of freedom, neutron-rich nuclei with neutron number $110<N<122$  show structural evolution. These isotopes show collective behavior and transit from the prolate shape in the ground state to the oblate shape as the neutron number approaches $N=126$. This shape transition was observed in Pt and Os isotopes \cite{ansari1986, wheldon2000, podolyak2009, nomura2011, john2014, nomura2018}. Furthermore, the wave functions of some of the excited states of the doubly magic nucleus such as $^{208}$Pb, and the nuclei around it have complex structures because they involve core breaking. It is a difficult task to study these states because of the increased model space, but at the same time, it is interesting to examine these nuclei to explore the collective nature of these states. It can be done via particle-hole excitations in the nucleus. In $^{208}$Pb, the first excited state, i.e., $3^-$ \cite{brown2000, isacker2022} shows collective behavior thus making the wave function more complex and mixed. In $^{207}$Tl also, the high spin states are studied by considering particle-hole excitations \cite{berry2020}. Further, Kumar \textit{et al.} \cite{kumar2021} studied $\beta$ decay $\log ft$ values for the decay of $^{207}$Hg into one-proton-hole nucleus $^{207}$Tl using particle-hole truncation and predicted some of the spin-parity states in $^{207}$Tl where experimentally assignment was tentative.

	Further, limited study of heavier neutron-rich nuclei is found because of their structure complexities via both theoretical and experimental aspects. Additionally, it is challenging to experimentally produce the nuclei in the vicinity of $^{208}$Pb. We have studied $\beta$-decay properties of neutron-rich nuclei in the $^{132}$Sn region \cite{sharma2023} and in the northeast region of $^{208}$Pb using shell-model calculations \cite{sharma2022,srivastava2023}. Thus, we intend to examine neutron-rich nuclei in the south region of $^{208}$Pb with the nuclear shell model.

 In the present work, we have systematically investigated the $\beta$-decay properties of the nuclei in the south region of $^{208}$Pb. To evaluate the $\beta$-decay rates, we have utilized the shell model using the slightly modified effective interaction by Yuan {\it et al.} \cite{Yuan2022}. With this interaction, they have systematically investigated the nuclear structure properties such as the energy spectra and electromagnetic properties, of nuclei lying in the lead region. Here, our interest is to investigate the $\beta$ decay properties of nuclei in the same region with this effective interaction. We have also suggested some spin-parity of the unconfirmed states from the experimental side based on the shell model calculated $\log ft$ values. In general, the present shell model predicted decay rates are found to be well consistent with the available experimental data.   
 %In the Ref. \cite{Yuan}, the systematic study of the modified interaction 
 %In the present work, we adopt a slightly modified effective interaction for nuclei in the south region of $^{208}$Pb, which can very well describe the energy spectra and electromagnetic properties of nuclei in the mentioned region \cite{Yuan2022}. 
 %Thus, our further motivation is to discuss this new effective interaction validity and predictive power in the case of weak decay processes. 
 The present study will add more information about the $\beta$ decay properties of nuclei located in south region of $^{208}$Pb, which  will be very useful for the upcoming experiments in the {\it Terra Incognita} region. 
	The present work is organized as follows.  In Sec. \ref{formalism}, we briefly explain the shell-model Hamiltonian and formalism of $\beta$ decay theory followed by the quenching factor. In Sec. \ref{result}, results and their explanation are given. $\beta$ decay properties like $\log ft$, shape factor values, half-lives, and partial decay rates have been calculated in this section. The discussion ends with a conclusion in the Sec. \ref{Conclusion}.

		\section{Formalism} \label{formalism}
	
	\subsection{Shell-model Hamiltonian}
	One of the significant complex tasks in the shell-model calculation is determining the effective Hamiltonian operator. To resolve this, one has to choose a suitable effective interaction that describes the nuclear properties in the desired mass region. In shell-model Hamiltonian construction, a particular model space is chosen, the single-particle energies and two-body interaction between the nucleons of the selected model space are determined. Then, the Hamiltonian is constructed and diagonalized to find the eigenvalues and eigenfunctions. The shell-model Hamiltonian \cite{suhonen2007} can be defined as
	\begin{equation}
	H = T + V = \sum_{\alpha}{\epsilon}_{\alpha} c^{\dagger}_{\alpha} c_{\alpha} + \frac{1}{4} \sum_{\alpha\beta \gamma \delta}v_{\alpha \beta \gamma \delta} c^{\dagger}_{\alpha} c^{\dagger}_{\beta} c_{\delta} c_{\gamma},
	\end{equation}
	% where $\alpha = \{n,l,j,t\}$ denotes the single-particle state with $n,l,j,t$ being principal, orbital, total angular momentum, and isospin quantum numbers, respectively.
 where $c^\dagger_\alpha$ and $c_{\alpha}$ denote creation and annihilation operators of the single-particle state $\alpha$, respectively. The quantity $\epsilon_{\alpha}$ corresponds to the single particle energy, and $v_{\alpha \beta \gamma \delta} = \langle\alpha \beta | V | \gamma \delta\rangle $ are the antisymmetrized two-body matrix elements.

	In the present work, the shell-model Hamiltonian consists of the model space with proton number $50 \leq Z \leq 82$ and neutron number $82  \leq N 
 \leq 184$. There are five proton orbitals (below $Z=82$)  and 13 neutron orbitals (six below $N=126$ and seven above $N=126$). The proton orbitals are $0g_{7/2}, 1d_{5/2}, 1d_{3/2}, 2s_{1/2}, 0h_{11/2}$ (PO5) and the neutron orbitals are $0h_{9/2}, 1f_{7/2}, 1f_{5/2}, 2p_{3/2}, 2p_{1/2}, 0i_{13/2}$ (NO6) and $0i_{11/2}, 1g_{9/2}, 1g_{7/2}, 2d_{5/2}, 2d_{3/2}, 3s_{1/2}, 0j_{15/2}$ (NO7). Further, in the case of particle-hole excitation, the six proton orbitals above $Z=82$, i.e., $0h_{9/2}, 1f_{7/2}, 1f_{5/2}, 2p_{3/2}, 2p_{1/2}, 0i_{13/2}$ are also considered. The Hamiltonian is constructed with the help of the KHHE interaction  \cite{warburton1991} for the proton-proton interaction in PO5,  the neutron-neutron interaction in NO6, and the proton-neutron interaction between PO5 and NO6 orbitals. The KHPE interaction \cite{warburton1991app} is also included for proton-proton interaction above $Z=82$ and neutron neutron interaction inside NO7 orbitals. Further, the monopole-based universal ($V_{MU}$) interaction  \cite{otsuka2010}, and spin-orbit interaction from the M3Y interaction \cite{bertsch1977}, which is referred as $V_{MU}$+LS, is considered for proton-neutron interaction between PO5 and NO7 and neutron-neutron interaction between NO6 and NO7 orbitals. Using the above-mentioned effective interaction, the nuclear wave functions have been calculated using shell-model code {\footnotesize KSHELL} \cite{shimizu2019}. 
 
 %It is also considered in the case of proton and neutron core excitations.

  It is very challenging to perform large-scale shell-model calculations in the entire Pb region without truncation. 
   In our study, we have performed two sets of calculations: (i) the model space is taken as $50 \leq Z \leq 82$ and $82  \leq N \leq 126$ for nuclei with $N\leq 126$
  and (ii) the two major shells are taken for the $N > 126$ nuclei.
  In case (i), we apply no truncation for this model space.
  % (i) case (i) which means no truncation for nuclei with $N\leq 126$ and (ii) particle-hole excitations for nuclei with $N > 126$. In the case of case (i) with $A\geq199$ and $N \leq 126$, we have performed calculation without any truncation in the model space $50 \leq Z \leq 82$ and $82  \leq N \leq 126$.
  %When we move towards the lower mass region, the dimension starts increasing with increasing hole numbers, making the shell-model calculations difficult. 
 % Thus, we have excluded lower mass nuclei in this work.
 % When the neutron number is $N>126$, 
 On the other hand, the beta decay of the nuclei with $N>126$ and $Z<82$ would make a neutron hole below the $N=126$ gap. We have extended our model space, i.e., $50 \leq Z \leq 126$ and $82 \leq N \leq 184$ for nuclei with neutron numbers above 126.  But the huge shell-model dimension prevents us from the calculation without truncation. 
 Thus, in case (ii), we apply the truncation based on the particle-hole excitations. We allow up to one proton and one neutron excitations across the $Z=82$ and $N=126$ shell closures simultaneously for these nuclei. 
 The truncation of one neutron and one proton excitation seems to be a good choice for beta decay in core-breaking nuclei ($Z<82$ and $N>126$) near $^{208}$Pb \cite{warburton1990} because the nuclear states at higher energies are due to core breaking. It is hard to study these states with increasing valence space due to the large dimension, but it can be done via particle-hole excitations across the shell closure. Further, there are many orbitals above this shell-closure that are involved in the first forbidden transition (see Fig. \ref{north}), for instance, transition of $\nu 1g_{9/2}$ to $\pi 0h_{9/2}$ orbital in $0^-$ transition.
 We have to consider contributions from different orbitals as shown in the  figure in order to quantify the effects of nuclear matrix elements for these transitions.

 \begin{figure}
 \includegraphics[width=90mm,height=95mm]{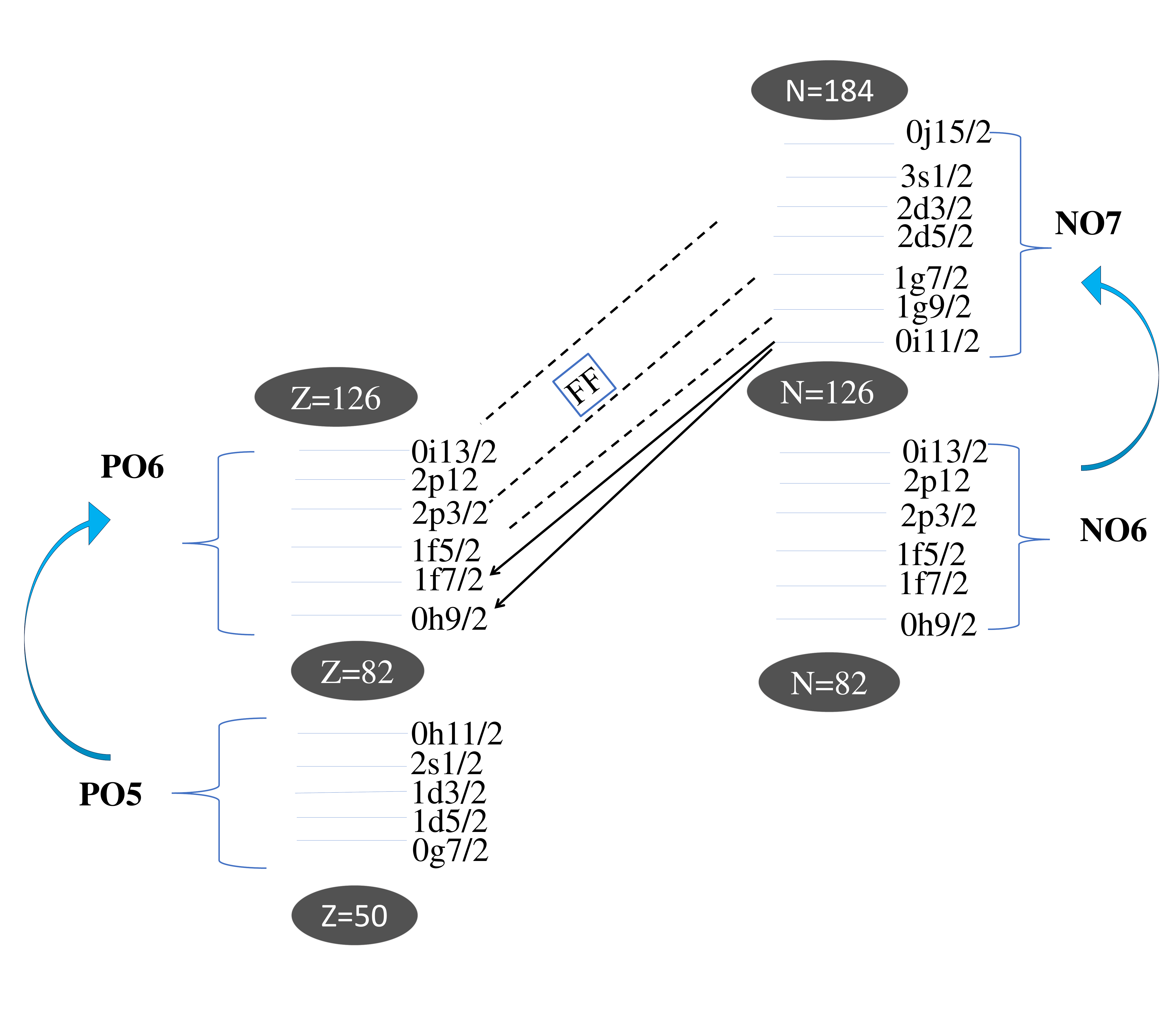}
			\label{north}
\vspace{-1.5cm}			\caption{Involvement of higher orbitals in first forbidden transitions for $N>126$ nuclei in $^{208}$Pb region. As an example, the first forbidden transitions have shown from $0i_{11/2}$ to $0h_{9/2}$, $1f_{7/2}$, orbitals in solid lines and similarly holds for all possible first-forbidden transitions in dashed lines.} \label{north}
		\end{figure}

	%\begin{widetext}

 \subsection{ Formalism for allowed and forbidden  $\beta$ decay transitions} \label{beta}
	In this study, we have computed the decay rate for both allowed and first-forbidden transitions. The detailed formalism of the beta decay is described in  Refs. \cite{behrens1982,schooper1966,suzuki2012,zhi2013}.
 %Behrens and Bh$\ddot{\rm u}$ring book \cite{behrens1982} (see also in \cite{schooper1966}).
 %The generalized formalism for allowed and first-forbidden beta decay can be found in Refs. \cite{suzuki2012,zhi2013}.
 The partial half-life of the $\beta$-decay transition between an initial state and a final state can be related to the phase space factor as
 \begin{eqnarray}
t_{1/2} = \frac{\kappa}{f}     
 \end{eqnarray}
 with constant $\kappa$ whose value is $\kappa = 6289~{\rm s}$ \cite{patrignami2016}. The form of the phase space factor $f$ is
 \begin{eqnarray}\label{eqn:phase}
     f = \int_{1}^{W_0}C(W)F_0(Z, W)\sqrt{(W^2-1)}W(W_0-W)^2dW. \nonumber \\
 \end{eqnarray}
     
Here, the term $C(W)$ is the shape factor that contains nuclear structure information of the transitions, $W$ is the energy of the emitted electron in units of the electron mass, and $W_0$ is the maximum energy of $W$. The factor $F_0(Z, W)$ is the Fermi function, which takes into account the Coulomb interaction between the daughter nucleus and emitted $\beta$ particles.

In the case of Gamow-Teller (GT) transitions, the shape factor remains independent of the emitted electron energy and can be expressed as
\begin{eqnarray}
    C(W) = B(GT) = {g_{\rm A}^2}\frac{\langle{f}||\sum_{\rm k}{\pmb{\sigma}^k{t}_-^k}||i\rangle^2}{2J_i+1},
\end{eqnarray}
where the reduced matrix element is computed solely with respect to the spin operator $\pmb{\sigma}$ and the $t_-$ refers to the isospin operator for the $\beta^-$ decay; we use the convention $t_-{|n\rangle}=|p\rangle$. The $g_{\rm A} = -1.2701  (25)$ \cite{Beringer} is the axial vector coupling constant, and $J_i$ denotes the angular momentum of the initial state.

In the case of the first-forbidden (FF) $\beta$ decay, $C(W)$ has the form
\begin{eqnarray}
    C(W) = k_0+k_1W+k_{-1}/W+k_2W^2,
\end{eqnarray}
where the coefficient ${k_n}$'s ($n= -1,~ 0,~ 1,~2  $) depends on the FF nuclear matrix elements, which can be written   
\begin{align}
{\rm for ~rank~ 0}~~ ~~~~~\nonumber\\
    k_{-1} =& -\frac{2}{3}\mu_1\gamma_1\zeta_0w \nonumber \\
    k_0 ~= &~ \zeta_0^2+\frac{1}{9}w^2,\\   
{\rm for~ rank~ 1}~~ ~~~~~\nonumber\\
    k_{-1} = &~ \frac{2}{3}\mu_1\gamma_1\zeta_1(x+u), \nonumber \\
    k_0 ~ = &~\zeta_1^2+\frac{1}{9}(x+u)^2-\frac{4}{9}\mu_1\gamma_1u(x+u) \nonumber \\
            &~+\frac{1}{18}W_0^2(2x+u)^2-\frac{1}{18}\lambda_2(2x-u)^2, \nonumber \\
    k_1~ = & ~-\frac{4}{3}uY-\frac{1}{9}W_0(4x^2+5u^2), \nonumber \\
    k_2~ = &~ \frac{1}{18}\left[8u^2+(2x+u)^2+\lambda_2(2x-u)^2\right],\\ 
~~~~{\rm and ~for~ rank~ 2}\nonumber \\
%\end{align}
%\begin{align}
    k_0~ = &~ \frac{1}{12}z^2(W_0^2-\lambda_2), \nonumber \\
    k_1~ =&~ -\frac{1}{6}z^2W_0, \nonumber \\
    k_2 ~= &~\frac{1}{12}z^2(1+\lambda_2),  
\end{align}
with 
\begin{align}
\zeta_0 = &~ V+\frac{1}{3}wW_0, ~~~~ ~~~V=~ \xi^\prime{v}+\xi{w}', \\
\zeta_1 = &~ Y+\frac{1}{3}(u-x)W_0, ~~~ Y = ~\xi^{\prime}y-\xi(u^{\prime}+x^{\prime}),
\end{align}
where $\xi$ and $\gamma_1$ are constants given by $\xi = \alpha{Z}/(2R)$ and $\gamma_1=\sqrt{1-(\alpha{Z})^2}$ with $\alpha$ the fine-structure constant and $R$ is the charge radius for a uniform charge distribution. In this work, approximated values of the quantities $\mu_1$ and $\lambda_2$ are used, i.e., $\mu_1\approx 1$ and $\lambda_2\approx 1$ \cite{warburton1988ml}. The quantities $w$, $\xi^\prime{v}$, $\xi^\prime{y}$, $x$, $u$, and $z$ along with their primed quantities $w^\prime$, $x^\prime$, and $u^\prime$ denote the matrix elements corresponding to FF operators, which can be expressed as \cite{behrens1982}
\begin{subequations}
\begin{align}
    w= &~  -g_{\rm A}\sqrt{3}\langle{f}||ir\left[{C}_1\times\pmb{\sigma}\right]^0{t}_-||i\rangle{C}, \\
    \xi^\prime{v} = &~  g_{\rm A}\sqrt{3}\langle{f}||(i/M_N)\left[\pmb{\sigma}\times{\nabla}\right]^0{t}_-||i\rangle{C}, \\
    %\xi^\prime{v} = &~ -g_{\rm A}\langle{f}||\gamma_1{t}_-||i\rangle\\
    \xi^\prime{y} = &~ -\langle{f}||(i/M_N)\nabla{t_-}||i\rangle{C},\\
    x =&~  -\langle{f}||ir{C}_1{t}_-||i\rangle{C} , \\
    u=&~ -g_{\rm A}\sqrt{2}\langle{f}||ir\left[{C}_1\times\pmb{\sigma}\right]^1t_-||i\rangle{C},\\
    z = &~ 2g_{\rm A}\langle{f}||ir\left[{C}_1\times\pmb{\sigma}\right]^2t_-||i\rangle{C},
\end{align}
\end{subequations}
with quantities $C_L = \sqrt{4\pi/(2L+1)}Y_L$,  $C=1/\sqrt{2J_i+1}$, and the prime matrix elements take into account the nuclear charge distribution, which is computed by multiplying the operators with the factor
\begin{eqnarray}
  \begin{cases}
 1-\frac{1}{5}\left(\frac{r}{R}\right)^2 & {\rm for}~ 0\leq r \leq R,\\
 \frac{R}{r}-\frac{1}{5}\left(\frac{R}{r}\right)^3 & {\rm for}~ R\leq r.
 \end{cases}
\end{eqnarray}

Based on the conserved vector current theory, the matrix element $\xi^\prime{y}$ can be expressed in terms of the $x$ matrix element \cite{behrens1982,warburton1988ml} 
\begin{eqnarray}
    \xi^\prime{y} = E_{\gamma}x,
\end{eqnarray}
where $E_\gamma$ is the  energy difference between the isobaric analog of the initial and final state, which can be defined as
\begin{eqnarray}
    E_\gamma = E_{\rm ias (i)}-E_f=Q+\Delta{E_C}-\delta{m},
\end{eqnarray}
where $\delta m = (m_n-m_p-m_e)c^2 = 0.782$ MeV, and $Q$ is the $\beta$ decay $Q$ value. In the present study,  we have used experimental $Q$ value from Ref. \cite{nndc}. $\Delta{E}_{C}$ represents the Coulomb displacement energy between isobaric analog states. Here, we have used an approximated value from \cite{antony1997}
\begin{eqnarray}
    \Delta{E_C}=1.4136(1)\Tilde{Z}/A^{1/3}-0.91338(11)~ {\rm MeV}
\end{eqnarray}
with $\Tilde{Z}=(Z_i+Z_f)/2$.

		\begin{figure*}

                \includegraphics[width=85mm]{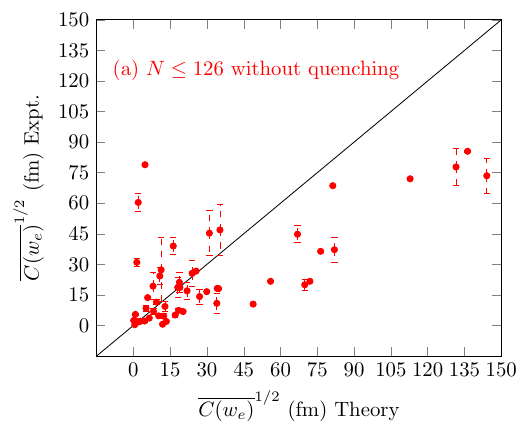}
			\includegraphics[width=85mm]{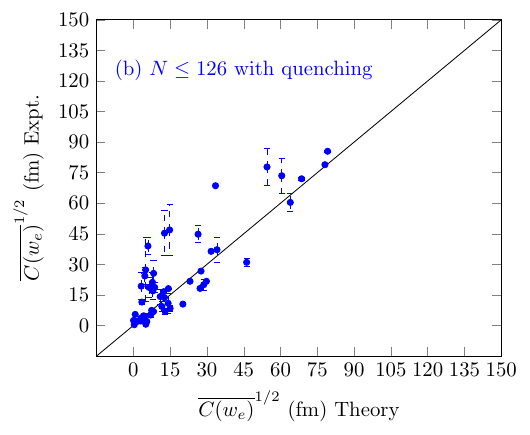}
      \includegraphics[width=85mm]{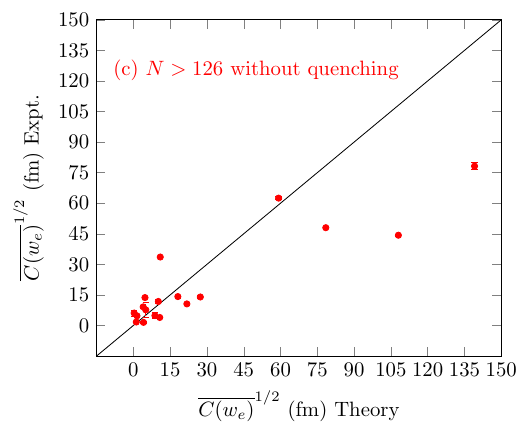}
			\includegraphics[width=85mm]{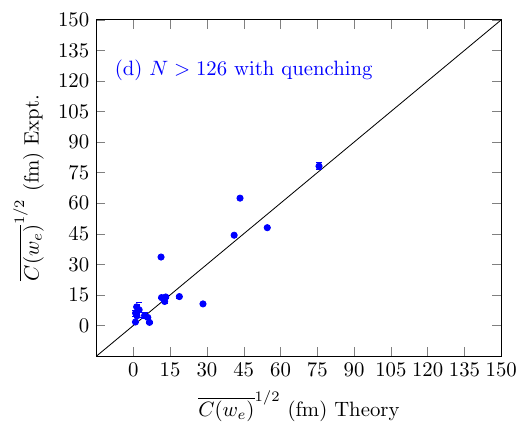}
			\label{shape1}
			\caption{Comparison of shell-model and experimental average
				shape factors. } \label{shape1}
		\end{figure*}

	\subsection{Quenching factor}
	In shell-model calculations, the Gamow-Teller strength gets overestimated, which is corrected by introducing a quenching factor in the Gamow-Teller operator \cite{martinez1996,brown1985}. The truncation of full Hilbert space into restricted model space, impulse approximation, and other nuclear medium effects significantly impact the Gamow-Teller strength \cite{suhonen2018,kostensalo2017}. This quenching factor $q_A$ in the GT operator can be calculated via the $\chi$-squared fitting method by comparing the shell-model and experimental $M(GT)$ (Gamow-Teller matrix element) values as shown in Ref. \cite{kumar2020}. By comparing with the Ikeda sum rule, the GT quenching factor in $^{90}$Zr was found to be $0.88\pm0.06$ from the total GT strength obtained by the $(p, n)$ and $(n, p)$ reactions on $^{90}$Zr over an excitation energy region of 0$-$70 MeV \cite{wakasa1997,yako2005}. The GT quenching factor for heavy nuclei such as $^{124}$Sn, and $^{208}$Pb was also found to be largely quenched by comparing the GT strength near the giant GT resonance region obtained from $(p, n)$ reactions and the Ikeda sum rule \cite{gaarde1981}. The quenching factor for $g_A$ was obtained by comparing the experimental $\beta$-decay strength and calculated shell-model strength within $pf$ shell and $sd$ shell. It was found to be $0.744\pm0.015$ \cite{martinez1996} and $0.77\pm0.02$ \cite{wildenthal1983} for the $pf$ shell and $sd$ shell cases, respectively. Further, the first forbidden transition strengths are also overestimated by shell-model calculations. Previous studies \cite{rydstrom1990, warburton1990, warburton1991, zhi2013} indicate the operators associated with ranks 0, 1, and 2 in the first-forbidden $\beta$ decay necessitate quenching factors. To obtain the quenching factors for each operator individually, we have minimized the $\chi$-squared function of the experimental and theoretical average shape factor, where experimental average shape factor values have been calculated from the experimental $\log ft$ values as shown in Ref. \cite{choudhary2021}. To make a comparison with experimental data, we can define a quantity referred to as the average shape factor
\begin{eqnarray}
    \overline{(C(W))} = f/f_0,
\end{eqnarray}
where, the quantity $f$ is taken from Eq. (\ref{eqn:phase}) and $f_0$ is

\begin{eqnarray}
f_0 = \int_1^{W_0} \sqrt{(W^2-1)}W(W_0-W)^2 F_0(Z,W)dW.
\end{eqnarray}

	Further, in the case of $\Delta J=0^-$ transitions, the matrix element of the time component of the axial-vector current, i.e., $\xi^\prime{v}$  gets enhanced over the impulse approximation with the help of mesonic enhancement factor \cite{towner1992,warburton1991}. 

\begin{table}[!h]
\vspace{-1ex}
\caption{\label{qfac_ff} Quenching factors adopted in the present study  for case (i) and case (ii) in comparison to Zhi {\it et al.} \cite{zhi2013} that define the effective operators. Please see text for more details.}
\begin{ruledtabular}
\begin{tabular}{rcccccc}
 & GT &\multicolumn{5}{c}{FF} \T\B \\
\cline{3-7}
  &  & $q_w$ & $q_{\xi^\prime{v}}$ & $q_x$ &$q_u$ & $q_z$ \T\B\\
\hline
Case (i) & 0.70 & 0.41  &  1.265  & 0.40     & 0.43  &  0.66  \T\\
Case (ii) & 0.70 & 0.57  & 1.320   & 0.54    & 0.30  & 0.54  \T\B\\
Ref. \cite{zhi2013} &  & 0.66  & 1.266   & 0.51    & 0.38  & 0.42  \T\B\\
\end{tabular}
\end{ruledtabular}
\end{table}

\begin{table*}
		\centering
		\label{table1}
\caption{Comparison between shell-model (SM) and experimental \cite{zhu2008, singh2007, kondev2005, chiara2010, kondev2023, kondev2020, kondev2008, kondev2007} $\log ft$ and average shape factor values for case (i) ($ N <  126 $) using $q_w = 0.41$, $q_{\xi^\prime{v}} = 1.265$, $q_x = 0.40$, $q_u = 0.43$, $q_z = 0.66$ in first forbidden transitions and $q_A=0.70$ for GT transitions. Here, double brackets represent the unconfirmed states. (‘NA’: ‘not available’.)}\label{table1}
\begin{ruledtabular}
			\begin{tabular}{lccccccccccc}
				\multicolumn{2}{c}{~~~~~Transition} &Energy& \multicolumn{2}{c}{$\log ft$} &\multicolumn{2}{c}{$ {[\overline{C(w_{e})}]}^{1/2}$} \T\B \\
				\cline{1-2}
				\cline{4-5}
				\cline{6-7}
				
				Initial ($J_i^\pi$)  & Final ($J_f^\pi$)  &(keV) &Expt.  &SM  & Expt. &SM \T\B\\\hline\T\B

				$^{202}$Ir($(1^-)$) \cite{zhu2008} &$^{202}$Pt($0_1^+$) &0.0 &$\geq$ 5.9&8.698 &$\leq$34.4 &1.37 \T \\ 
				&$^{202}$Pt($(2_1^+)$)  &534.90(20) & NA&7.574  & NA &5.00 \\
				%&$^{202}$Pt($(4_1^+)$)&3rd FU &1253.6(3) & NA& 12.635 & NA &2198.315 \\
				
				$^{202}$Ir($(2^-)$) \cite{zhu2008} &$^{202}$Pt($0_1^+$) &0.0 &$\geq$ 5.9&9.149 &$\leq$34.4 &0.816\\
				&$^{202}$Pt($(2_1^+)$) &534.90(20) & NA&8.665  & NA &1.42 \\
				&$^{202}$Pt($(4_1^+)$)&1253.6(3) &NA & 8.570 & NA &1.59 \\
				
				$^{199}$Pt($5/2^-$) \cite{singh2007} &$^{199}$Au($3/2_1^+$) &0.0&6.30(1) &6.246 &21.7(2) &23.1\\
				&$^{199}$Au($1/2_1^+$) &77.170(21)&  $>$8.0& 10.438 &$<$3.06 &0.185\\
				&$^{199}$Au($5/2_1^+$) &317.174(24) &7.30(1)&7.140 &6.86(8) &8.24\\
				&$^{199}$Au($3/2_2^+$) &323.605(25) &7.62(3)&7.736 &4.74(16) &4.15\\
				&$^{199}$Au($(7/2_1)^+$) &493.76(3) &8.03(9)& 8.068& 2.96(31) &2.83\\
				&$^{199}$Au($5/2_2^+$) &542.884(23) &6.45(1)&6.104 &18.2(2) &27.2\\
				&$^{199}$Au($7/2_1^-$) &734.64(3)&  6.45(1)&5.941 &0.0472(54) &0.0849\\
				&$^{199}$Au($3/2_2^+$) &791.760(25) &6.53(1)& 6.784& 16.6(2)& 12.4\\
				&$^{199}$Au($1/2_2^+$) &822.7(3) &8.18(2) &11.062 &2.49(6) &0.0902\\
				&$^{199}$Au($(3/2_3^+)$) &968.29(4) &6.82(1)& 6.788&11.9(1) &12.4\\
				&$^{199}$Au($(5/2_3^+)$) &968.29(4) &6.82(1)&7.339 &11.9(1) &6.55\\
				&$^{199}$Au($(3/2_4^+)$) &1070.02(14)&  7.77(2)& 7.576& 3.99(9)& 4.99\\
				&$^{199}$Au($(5/2_4^+)$) &1070.02(14)&  7.77(2)&7.022 &3.99(9) & 9.44\\
				&$^{199}$Au($(3/2_5^+)$) &1103.99(13)&  7.64(2)&9.386 &4.64(11) & 0.621\\
				&$^{199}$Au($(5/2_5^+)$) &1103.99(13)&  7.64(2)&7.600 &4.64(11) & 4.85\\
				&$^{199}$Au($(7/2_2^+)$) &1103.99(13)&  7.64(2)& 7.969& 4.64(11)& 3.17\\
				&$^{199}$Au($(3/2_6^+)$) &1159.01(7)&  7.46(2)&9.621& 5.70(13) & 0.474\\
				&$^{199}$Au($(5/2_6^+)$) &1159.01(7)&  7.46(2)& 7.289& 5.70(13) & 6.94\\
				&$^{199}$Au($(7/2_3^+)$) &1159.01(7)&  7.46(2)&7.808 & 5.70(13) & 3.82\\
				&$^{199}$Au($(3/2_7^+)$) &1249.4(3)&  8.16(3)&8.691 & 2.55(9) & 1.38\\
				&$^{199}$Au($(5/2_7^+)$) &1249.4(3)&  8.16(3)& 7.470& 2.55(9) & 5.64\\
				&$^{199}$Au($(7/2_4^+)$)&1249.4(3)&  8.16(3)& 8.252 & 2.55(9) & 2.29\\
				&$^{199}$Au($(5/2_8^+)$) &1396.22(19)&  7.07(2)&7.347 & 8.93(21) & 6.49\\
				&$^{199}$Au($(7/2_5^+)$)&1396.22(19)&  7.07(2)& 8.192 & 8.93(21) & 2.46\\

				$^{202}$Pt($0^+$) \cite{zhu2008} &$^{202}$Au($(1_1^-)$)&0.0 &$\geq$8.2& 5.341 &$\leq$2.43 &65.4\\
				
				$^{199}$Au($3/2^+$) \cite{singh2007} &$^{199}$Hg($1/2_1^-$)&0.0 &7.50(9)&9.209&5.45(56)&0.761\\
				&$^{199}$Hg($5/2_1^-$)&158.37859(10) &5.850(9)&5.969&36.4(4) &31.7\\
				&$^{199}$Hg($3/2_1^-$) & 208.20494(10) &6.118(9)&6.091& 26.7(3)& 27.6\\

				$^{200}$Au($(1^-)$) \cite{kondev2007} &$^{200}$Hg($0_1^+$) &0.0 &6.93(5)&6.360&10.5(6)&20.2\\
				&$^{200}$Hg($2_1^+$)&367.943(10) &7.85(10)&7.980&3.64(42)&3.13\\
				&$^{200}$Hg($0_2^+$)& 1029.344(17) &8.46(10)&9.085&1.80(21) &0.878\\
				&$^{200}$Hg($2_2^+$)& 1254.098(17) &8.37(11)&7.491&2.00(25)& 5.50\\
				&$^{200}$Hg($0_3^+$)& 1515.173(18) &NA &7.370&NA &6.32\\
				&$^{200}$Hg($1_1^+$)& 1570.275(17) &7.12(14)&6.620&8.43(136)& 15.0\\
				&$^{200}$Hg($2_3^+$)&1573.665(18) &NA &9.754&NA &0.406\\
				&$^{200}$Hg($2_4^+$)& 1593.434(18) &5.83(14)&5.906&37.2(60)& 34.1\\
				&$^{200}$Hg($1_2^+$)&1630.892(17) &6.20(15)&7.630&24.3(42)& 4.69\\
				&$^{200}$Hg($2_5^+$)& 1641.443(17) &7.55(16)&7.273&5.14(95)& 7.07\\
				&$^{200}$Hg($1_3^+$) & 1718.304(17)&7.29(18)&6.747&6.94(144)& 13.0\\
				&$^{200}$Hg($2_6^+$) &1730.927(17)& 8.37(24)&9.462&2.00(55)& 0.569\\
    	\end{tabular}
		\end{ruledtabular}
	\end{table*}
	\addtocounter{table}{-1}
	\begin{table*}
		\caption{{\it (Continued.)}}
		% 	
		% 	%\vspace*{5mm}
		\begin{ruledtabular}
			
			\begin{tabular}{lccccccccccc}
				
				\multicolumn{2}{c}{~~~~~Transition} &Energy& \multicolumn{2}{c}{$\log ft$} &\multicolumn{2}{c}{$ {[\overline{C(w_{e})}]}^{1/2}$} \T\B \\
				\cline{1-2}
				\cline{4-5}
				\cline{6-7}
				
				Initial ($J_i^\pi$)  & Final ($J_f^\pi$)  &(keV) &Expt.  &SM  & Expt. &SM \T\B\\\hline\T\B 

				&$^{200}$Hg($0_4^+$)&1856.779(18) &9.38(21)&7.559&0.625(151)&5.09\\
				&$^{200}$Hg($2_7^+$)&1882.859(17) &6.67(23)&6.891&14.2(37)& 11.0\\
				&$^{200}$Hg($2_8^+$)&1972.280(18) &6.4(3)&7.958&19.3(67)& 3.21\\
				&$^{200}$Hg($1_4^+$)&2061.256(18) &6.1(5)&7.571&27.3(157)& 5.02\\ 
				
					$^{200}$Au$^{\rm m}(12^-$) \cite{kondev2007} &$^{200}$Hg($11_1^-$) & 2641.60(21)& 6.1(3)& 5.938&0.0707(244)&0.0852\\

				$^{201}$Au$(3/2^+$) \cite{kondev2023} &$^{201}$Hg($3/2_1^-$) & 0.0& $\geq$5.717&5.955 &$\leq$42.4 &32.3 \\
				&$^{201}$Hg($1/2_1^-$) & 1.5648(10)&NA &9.123 &NA &0.841 \\
				&$^{201}$Hg($5/2_1^-$) & 26.2738(3)& NA& 5.789&NA & 39.0\\
				&$^{201}$Hg($3/2_2^-$) & 32.168(14)& $\approx$6.7&6.767 &$\approx$13.7 &12.7 \\
				&$^{201}$Hg($1/2_2^-$) & 167.48(4)& 6.85(8)& 7.891&11.5(11) & 3.47\\
				&$^{201}$Hg($(5/2_2)^-$) & 384.605(17)& 7.07(8)& 7.222& 8.93(82)&7.50\\
				&$^{201}$Hg($5/2_3^-$) & 464.41(4)& 7.23(11)& 7.225& 7.43(94)& 7.47\\

				$^{202}$Au$((1^-)$) \cite{zhu2008} &$^{202}$Hg($0_1^+$) &0.0&$\sim$5.3& 5.921 & $\sim$68.6 & 33.5\\
				&$^{202}$Hg($2_1^+$) &439.512(8) &NA &7.008&NA &9.60\\
				&$^{202}$Hg($2_2^+$) &959.92(5) &6.9(4)& 6.666 & 10.9(50) & 14.2\\
				&$^{202}$Hg($(1_1^+)$) &1347.92(7) &5.80(19)&6.263 &38.6(84) &22.6\\
				&$^{202}$Hg($(2_3^+)$) &1347.92(7) &5.80(19)&6.159 &38.6(84) & 25.5\\
				&$^{202}$Hg($0_2^+$) &1411.37(12) &6.32(20)&7.187 &21.2(49) & 7.81\\
				&$^{202}$Hg($0_3^+$) &1564.78(8) &5.66(21)& 6.764 & 45.3(110) & 12.7\\
				&$^{202}$Hg($0_4^+$)&1643.62(10) &5.63(23)&6.633 & 46.9(124) & 14.8\\
				&$^{202}$Hg($(1_2^+)$) &1746.11(7) &5.35(24)& 7.175&64.7(179) &7.92\\
				&$^{202}$Hg($(1_1^-)$) &1746.11(7) &5.35(24)&8.846 & 0.168(46)&0.00299\\
				&$^{202}$Hg($(2_4^+)$) &1746.11(7) &5.35(24)&6.175 &64.7(179) &25.0\\
				&$^{202}$Hg($2_5^+$) &1852.14(9) &NA & 7.560&NA &5.08\\ 
				&$^{202}$Hg($(1_3^+)$) &1959.38(20) & NA& 5.961&NA &32.0 \\
				&$^{202}$Hg($(1_2^-)$) &1959.38(20) & NA&9.942 &NA &0.000848\\
				&$^{202}$Hg($(2_6^+)$) &1959.38(20) & NA&7.146 &NA &8.19\\
				
				$^{203}$Au($3/2^+$) \cite{kondev2005}&$^{203}$Hg($5/2_1^-$) &0.0 &5.19(10) & 5.500 & 77.8(90) &54.5\\
				&$^{203}$Hg($(1/2_1^-)$) &7.3(6) & NA&8.660 & NA& 1.43\\
				&$^{203}$Hg($(3/2_1^-)$) &50.8(4) &5.63(10)& 5.640 & 46.9(54) & 46.4\\
				&$^{203}$Hg($(3/2_2^-)$) &224.9(6)&  6.61(11)&6.867& 15.2(19)& 11.3\\
				&$^{203}$Hg($(1/2_2)^-$) &368.9(3) &6.45(13)&7.398 & 18.2(27) & 6.12\\
				&$^{203}$Hg($ (3/2_3)^-$) &368.9(3) &6.45(13)&6.028 &18.2(27) & 29.7\\
				&$^{203}$Hg($ (5/2_2)^-$) &368.9(3) &6.45(13)&7.567 &18.2(27) &5.04\\

				$^{204}$Au($(2^-)$) \cite{chiara2010} &$^{204}$Hg($0_1^+$) & 0.0 &$>$6.9&12.757&$<$10.9&0.0128\\
				&$^{204}$Hg($2_1^+$) & 436.552(8)&$>$7.4&6.815&$<$6.11& 12.0\\
				&$^{204}$Hg($4_1^+$)& 1128.23(11) &$>$7.5&8.657&$<$5.45& 1.44\\
				&$^{204}$Hg($(2_1^-)$) & 1828.71(11) &$\sim$5.5&8.804&$\sim$0.141& 0.00314\\
				&$^{204}$Hg($1_1^+$)& 1841.38(11) &$\sim$6.3&6.023&$\sim$21.7& 29.8\\
				&$^{204}$Hg($(2_2)^+$)& 1851.26(10) &$\sim$5.3&5.318&$\sim$68.6& 67.2\\
				&$^{204}$Hg($(3_1)^+$)& 1851.26(10) &$\sim$5.3&5.466&$\sim$68.6& 56.6\\
				&$^{204}$Hg($2_3^+$)& 1947.69(11) &$\sim$5.3&6.931&$\sim$68.6& 10.5\\
				&$^{204}$Hg($(2_4^+)$) & 1989.36(10) &$\sim$6.0&5.973& $\sim$30.6 & 31.6\\
				&$^{204}$Hg($2_5^+$)& 2088.51(10) &$\sim$6.6&7.364&$\sim$15.3 &6.3704\\
				&$^{204}$Hg($2_6^+$)& 2117.0(5) &$\sim$7.3&7.121&$\sim$6.86& 8.42\\
				
				%&$^{204}$Hg($3^-$)&Allowed &2812.83(24)& 0.66(7)&$\sim$6.0\\

				$^{203}$Hg($5/2^-$) \cite{kondev2005}&$^{203}$Tl($1/2_1^+$) &0.0 &$>$12.2&12.233&$<$0.0243&0.0234\\
				&$^{203}$Tl($3/2_1^+$)& 129.1955(10) &6.457(8)&6.661&18.1(2)& 14.3\\

    \end{tabular}
		\end{ruledtabular}
	\end{table*}
	\addtocounter{table}{-1}
	\begin{table*}
		\caption{{\it (Continued.)}}
		% 	
		% 	%\vspace*{5mm}
		\begin{ruledtabular}
			
			\begin{tabular}{lccccccccccc}
				
				\multicolumn{2}{c}{~~~~~Transition} &Energy& \multicolumn{2}{c}{$\log ft$} &\multicolumn{2}{c}{$ {[\overline{C(w_{e})}]}^{1/2}$} \T\B \\
				\cline{1-2}
				\cline{4-5}
				\cline{6-7}
				
				Initial ($J_i^\pi$)  & Final ($J_f^\pi$)  &(keV) &Expt.  &SM  & Expt. &SM \T\B\\\hline\T\B

    	$^{205}$Hg($1/2^-$) \cite{kondev2020} & $^{205}$Tl($1/2_1^+$) & 0.0 &5.257(11)&5.299 &72.0(9)&68.6\\
				&$^{205}$Tl($3/2_1^+$) & 203.65(19)&6.51(21)&7.169&17.0(41)& 7.97\\
				&$^{205}$Tl($5/2_1^+$) & 619.3(3) &8.28(22)&8.012&2.22(56)& 3.02\\
				&$^{205}$Tl($3/2_2^+$)& 1140.6(3) &7.65(22)&7.642&4.58(116)& 4.62\\
				&$^{205}$Tl($1/2_2^+$) & 1218.6(4) &7.03(25)&6.843&9.36(269)& 11.6\\
				&$^{205}$Tl($3/2_3^+$)& 1340.3(5) &6.43(22)&7.359&18.7(47)& 6.41\\
				&$^{205}$Tl($(1/2_3^+)$)& 1434.0(5) &5.6(3)&5.859&48.5(168)& 36.0\\
				
			$^{204}$Tl($2^-$) \cite{chiara2010} & $^{204}$Pb($0_1^+$) & 0.0 & 9.7839(15)& 9.791&0.393(1)& 0.390\\

				$^{206}$Tl($0^-$) \cite{kondev2008}&$^{206}$Pb($0_1^+$)&0.0 &5.1775(13)&5.187& 78.9(1)& 78.1\\
				&$^{206}$Pb($2_1^+$)&803.049(25) &8.32(3)&8.405& 2.12(7)& 1.92\\
				&$^{206}$Pb($0_2^+$) &1166.4(5) &5.99(6)&5.642&31.0(21)& 46.2\\ 
		\end{tabular}
\end{ruledtabular}
\end{table*}

	\begin{table*}
		\centering
		\label{table2}
		\caption{Comparison between shell-model (SM) and experimental \cite{ martin2007, chen2015, berry2020} $\log ft$ and average shape factor values for case (ii) using $q_w = 0.57$,  $q_{\xi^\prime{v}} = 1.320$,  $q_x = 0.54$,  $q_u = 0.30$,  $q_z = 0.54$ in first forbidden transitions and $q_A=0.70$ for GT transitions. Here, double brackets represent the unconfirmed states.}\label{table2}
		\begin{ruledtabular}
			\begin{tabular}{lccccccccccc}
				\multicolumn{2}{c}{~~~~~Transition} &Energy& \multicolumn{2}{c}{$\log ft$} &\multicolumn{2}{c}{$ {[\overline{C(w_{e})}]}^{1/2}$} \T\B \\
				\cline{1-2}
				\cline{4-5}
				\cline{6-7}
				
				Initial ($J_i^\pi$)  & Final ($J_f^\pi$)  &(keV) &Expt.  &SM  & Expt. &SM \T\B\\\hline\T\B 
				
					$^{207}$Hg($9/2^+$) \cite{berry2020}&$^{207}$Tl($11/2_1^-$) &1348.3(2) &7.2(4)& 8.221 & 7.69(354)&2.37 \\
				&$^{207}$Tl($7/2_1^-$) &2676.0(2) &7.8(3)&7.651 & 3.86(133) &4.58\\
				&$^{207}$Tl($5/2_1^-$) &2709.3(6) &$>$8.7&9.291& $<$1.37 &0.693\\
				&$^{207}$Tl($(9/2_1^-)$) &2912.6(3) &6.3(2)&5.384 & 21.7(50) &62.2\\
				&$^{207}$Tl($(9/2_2^-)$) &2985.8(3) &5.42(7)&6.237 & 59.7(48) &23.3\\
				&$^{207}$Tl($(7/2_2^-)$) &3013.8(3) &7.7(2)&6.397 & 4.33(100) &19.4\\
				&$^{207}$Tl($(9/2_3^-)$) &3104.9(3) &5.58(8) &6.048 & 49.7(46) &29.0\\
				&$^{207}$Tl($(9/2_4^-)$) &3143.2(3) &5.95(7)&5.731 & 32.4(26) &41.7\\
				&$^{207}$Tl($(5/2_2^-)$) &3197.3(5) &$>$9.5 &10.132 & $<$0.545 &0.263\\
				&$^{207}$Tl($(7/2_3^-)$) &3273.5(2) &6.34(8)&6.616 & 20.7(19) &15.1\\
				&$^{207}$Tl($(9/2_5^-)$) &3296.2(3) &6.17(8) &5.786 & 25.2(23) &39.2\\
				&$^{207}$Tl($(9/2_6^-)$) &3336.5(2) &5.81(7) &7.546 &38.1(31)  &5.16\\
				&$^{207}$Tl($(9/2_7^-)$) &3358.7(2) &6.01(7)&6.315 & 30.3(24) &21.3\\
				&$^{207}$Tl($(7/2_4^-)$) &3430.5(2) &6.65(8) &6.925 & 14.5(13) &10.6\\
				&$^{207}$Tl($(5/2_3^-)$) &3493.6(5) &8.63(9)&9.380 & 1.48(15) &0.625\\
				&$^{207}$Tl($(7/2_1^+)$) &3493.6(5) &8.63(9) &9.150 &0.00384(40)  & 0.00211\\
				&$^{207}$Tl($(7/2_5^-)$) &3493.6(5) &8.63(9) &6.302 &1.48(15) & 21.6\\
				&$^{207}$Tl($(11/2_1^+)$) &3569.7(4) &7.21(10) &10.164 & 0.0197(23) &0.000657 \\
				&$^{207}$Tl($(11/2_2^-)$) &3569.7(4) &7.21(10) &6.085 &  7.60(88)&27.8 \\
				&$^{207}$Tl($(9/2_8^-)$) &3581.3(2) &6.97(8)&7.766 &10.0(92)  &4.01\\
				&$^{207}$Tl($(7/2_6^-)$) &3592.4(4) &7.11(9) &7.232 & 8.53(88) & 7.41\\
				&$^{207}$Tl($(11/2_3^-)$) &3633.6(3) &6.34(8) &6.025 & 20.7(19) &29.8 \\
				&$^{207}$Tl($(11/2_4^-)$) &3644.2(3) &6.72(9) &7.722& 13.4(14) &4.22\\

			$^{208}$Tl($5^+$) \cite{martin2007}&$^{208}$Pb($5_1^-$) &3197.717(11) &5.61(1)&5.498 & 48.0(5)&54.6\\
				&$^{208}$Pb($4_1^-$) &3475.088(11) &5.68(1)&5.744 &44.3(5) & 41.1\\
				&$^{208}$Pb($5_2^-$) &3708.41(7) &5.38(1)& 5.695&62.5(7) & 43.5\\
				&$^{208}$Pb($6_1^-$) &3919.78(10) &6.68(4)& 6.732&14.0(6) &13.2\\
				&$^{208}$Pb($4_2^-$) &3946.42(20) &7.78(3)& 7.431&3.95(14) &5.90\\
				&$^{208}$Pb($5_3^-$) &3960.93(7) &5.92(1)&6.865 &33.6(4) & 11.3\\
				&$^{208}$Pb($4_3^-$) &3995.6(5) &8.5(2)& 9.101 &1.72(40) & 0.862\\
				&$^{208}$Pb($5_4^-$) &4125.28(17) &6.92(6)& 6.065 &10.6(7) & 28.4\\
				&$^{208}$Pb($5_5^-$) &4180.38(17) &6.70(2)& 6.850&13.7(3) & 11.5\\
				&$^{208}$Pb($4_4^-$) &4262.0(7) &8.6(2)& 7.334 &1.53(35) & 6.59\\
				&$^{208}$Pb($5_6^-$) &4296.28(20) &6.83(9)& 6.759&11.8(12) & 12.8\\
				&$^{208}$Pb($4_1^+$) &4323.4(4) &8.1(2)& 9.320&0.00707(163) &0.00173\\
				&$^{208}$Pb($4_5^-$) &4358.44(17) &7.05(5)& 8.710 & 9.14(53)& 1.35\\
				&$^{208}$Pb($6_2^-$) &4382.9(3) &7.4(2)& 9.053 &6.11(141) &0.911\\
				&$^{208}$Pb($6_3^-$) &4480.5(3) &6.67(5)& 6.425 &14.2(8) &18.8\\

				$^{209}$Tl($1/2^+$)	\cite{chen2015}&$^{209}$Pb($1/2_1^-$) &2149.42(4) &5.186(20)&5.213&78.2(18) &75.8\\
				&$^{209}$Pb($3/2_1^-$) &2315.90(16) &7.62(14)&8.653 &4.74(76) &1.44\\
				&$^{209}$Pb($(5/2_1^-)$) &2460.9(3) &8.38(23)& 10.398 & 1.98(52)& 0.194\\
				&$^{209}$Pb($3/2_3^-$) &2905.27(25) &6.48(6)&6.119 & 17.6(12) &26.7\\
				&$^{209}$Pb($3/2_4^-$) &3069.92(16) &6.255(21)&8.842 & 22.8(6) &1.16\\

			\end{tabular}
		\end{ruledtabular}
	\end{table*}

 \section{Results and Discussion} \label{result}
	In this work, the $\log ft$ values for allowed and first forbidden transitions corresponding to the nuclei in the south region of $^{208}$Pb are calculated. These nuclei include isotopes of Ir$\rightarrow$Pt, Pt$\rightarrow$Au, Au$\rightarrow$Hg, Hg$\rightarrow$Tl, Tl$\rightarrow$Pb, and Os$\rightarrow$Ir ($N=126$ only) transitions. At first, the $\log ft$ values for the mentioned nuclei have been calculated without using any quenching factor in the $\beta$ decay operators. Using these $\log ft$ values, the average shape factor values are calculated and compared with the experimental average shape factor values as shown in Figs. \ref{shape1}(a) and \ref{shape1}(c).  In this work, the experimental $Q$ values for the ground state (g.s.) and excited state are taken for the calculation of $\beta$-decay properties. Further, the quenching factors are calculated in the nuclear matrix elements of the operators of the first forbidden transitions. These values of quenching factors have been computed for both sets of nuclei, i.e., case (i) and case (ii),  and the obtained quenching factors for first forbidden transitions are 
 shown in Table \ref{qfac_ff}. 
 %$q_w = 0.41$,  $q_{\xi^\prime{v}} = 1.265$,  $q_x = 0.40$,  $q_u = 0.43$,  $q_z = 0.66$ for full-set and $q_w = 0.57$,  $q_{\xi^\prime{v}} = 1.320$,  $q_x = 0.54$,  $q_u = 0.30$,  $q_z = 0.54$ for $N>126$ set.} 
 The quenching factor values in $\xi^\prime{v}$ are larger than other quenching factors, which shows the enhancement of $\xi^\prime{v}$ matrix element compared to others. The calculated average shape factor values via the inclusion of the quenching factors are also compared with the experimental data as shown in Figs. \ref{shape1}(b) and \ref{shape1}(d). There are also GT transitions involved in the $\beta$ decay of these nuclei. Evaluating the quenching factor for these GT transitions is impractical as there are not enough confirmed GT transitions available for the mentioned nuclei in this work. Thus, the standard value of the quenching factor in axial-vector coupling constant $q_A=0.70$ for GT transitions is opted as mentioned in Refs. \cite{zhi2013,caurier2005,suzuki2012}.

	Now, using these values of quenching factors in allowed and first forbidden transitions, the $\log ft$ values for the mentioned nuclei are calculated and shown in Tables \ref{table1}, \ref{table2}, and \ref{table4}. We have used $\log f_0t$ values for all transitions, including unique transitions. In case the $\log f_{1u} t$ value is given experimentally for unique transitions,  we have converted those $\log f_{1u}t$ values to $\log f_0t$ values and further compared the experimental and shell-model $\log ft$ results. The discussion for the $\log ft$ values
 for different chains is given below.

 \subsection{ The {\textbf{ $\log ft$}} results}

\subsubsection{\bf{Ir$\rightarrow$Pt}}
	For Ir$\rightarrow$Pt chain, the $\log ft$ values corresponding to $^{202}$Ir are computed.  Experimentally, two spin-parity states, i.e., $(1^-)$ and $(2^-)$ are tentatively assigned for the ground state of $^{202}$Ir isotope. Thus, the shell-model $\log ft$ values corresponding to both spin-parity states have been calculated, and the results are shown in Table \ref{table1}. Assigning one particular spin-parity state to the ground state of $^{202}$Ir from the corresponding $\log ft$ values is difficult. Alternatively, by analyzing the half-life values (reported later in the text), one can interpret that $(1^-)$ is more suitable for the ground state of $^{202}$Ir. 
 
	\subsubsection {\bf{Pt$\rightarrow$Au}}
	Moving forward towards Pt$\rightarrow$Au chain, the $\log ft$ values corresponding to $\beta^-$ decay of $^{199,202}$Pt isotopes have been computed. For $^{199}$Pt($5/2^-$)$\rightarrow$ $^{199}$Au($3/2^+_1$) transition,  the shell-model $\log ft$ value is 6.246 which is close to the experimental $\log ft$ value, i.e., 6.30(1). Similarly, for $^{199}$Pt($5/2^-$)$\rightarrow$ $^{199}$Au($3/2^+_2$) transition, the shell-model $\log ft$ value is 7.736 and the experimental $\log ft$ value is 7.62(3). Overall, the shell-model and experimental $\log ft$ values are close to each other. Further, in the $^{199}$Au isotope, more than one spin-parity states are tentatively assigned experimentally at some energy levels. Thus, $\log ft$ values for all possible spin-parity states are calculated using shell-model calculations and compared with the corresponding experimental data.  For instance, at energy level 968.29(4) keV,  the shell-model $\log ft$ value (6.788) for $(3/2_3^+)$ state is closer to the experimental $\log ft$ value (6.82(1)) as compared to the other one. Thus, the $(3/2_3^+)$ state can be proposed as the assigned spin-parity state at 968.29(4) keV energy level. Similarly, at 1070.02(14) keV energy level, the  $(3/2_4^+)$ state can be proposed as the confirmed spin-parity state since its shell-model $\log ft$ value (7.576) is close to the experimental $\log ft$ value (7.77(2)). Further, at energy level 1103.99(13) keV and 1159.01(7) keV, the $(5/2^+_5)$ and $(5/2^+_6)$ can be proposed as the confirmed spin-parity states, respectively. In a similar manner, the $(7/2^+_4)$ and $(5/2^+_8)$ can be proposed as the confirmed spin-parity states at 1249.4(3) and 1396.22(19) keV energy levels, respectively. Moving forward, for $^{202}$Pt isotope, the shell-model calculated $\log ft$ value is very low compared to the experimental $\log ft$ value.
  %This is because we have only considered one transition for $^{202}$Pt, and experimentally spin-parity of the daughter nucleus is not yet confirmed.
 %Also, there are some forbidden unique transitions in the Pt chain for which SM $\log ft$ values are very high compared to the experimental $\log ft$ values.  
	\subsubsection{\bf {Au$\rightarrow$Hg}}
	In the case of Au$\rightarrow$Hg chain, the $\log ft$ values corresponding to $\beta^-$ decay of $^{199-204}$Au and $^{200}$Au$^{\rm m}$ isotopes have been calculated. For $^{199}$Au$(3/2^+) \rightarrow  ^{199}$Hg($5/2_1^-$) transition, 
 the shell-model $\log ft$ value is 5.969 which is close to the experimental $\log ft$ value, i.e., 5.850(9). Similarly, for $^{199}$Au$(3/2^+) \rightarrow 
 ^{199}$Hg($3/2_1^-$) transition, the shell-model $\log ft$ value is 6.091 and the experimental $\log ft$ value is 6.118(9). The shell-model $\log ft$ values corresponding to beta-decay of $^{200}$Au and $^{201}$Au are also good.  For instance, the shell-model $\log ft$ value corresponding to $^{200}$Au$((1^-)) \rightarrow ^{200}$Hg($2_4^+$) transition at 1593.434(18) keV energy level is 5.906 which is close to the experimental value, i.e., 5.83(14). Moreover, we have also computed $\log ft$ values in those cases where experimental data are unavailable. Further, in $^{202}$Hg isotope, three spin-parity states are tentatively assigned experimentally at 1746.11(7) keV energy level. Thus, comparing the shell-model $\log ft$ values with experimental data, the $(2_4^+)$ spin-parity state can be suggested at this energy level for $^{202}$Hg isotope. Moving forward, in the $\beta$ decay of $^{203}$Au isotope, all of the calculated shell-model $\log ft$ values agree with the corresponding experimental $\log ft$ values. 
  For instance, for $^{203}$Au($3/2^+$)$\rightarrow$ $^{203}$Hg($3/2^-_1$) transition, the shell-model $\log ft$ value is 5.640 which is very close to the experimental $\log ft$ value, i.e., 5.63(10). Similarly, for $^{203}$Au($3/2^+$)$\rightarrow$ $^{203}$Hg(($3/2^-_2$)) transition,  the shell-model $\log ft$ value is 6.867 and the experimental $\log ft$ value is 6.61(11). The shell-model $\log ft$ value for $^{203}$Au($3/2^+$)$\rightarrow$ $^{203}$Hg(($1/2^-_1$)) transition has been computed, which comes out to be 8.660. Further, three spin-parity states are tentatively assigned experimentally for $^{203}$Hg isotope at 368.9(3) keV energy level. The shell-model $\log ft$ values for all these three spin-parity states are computed and compared with the experimental data. Based on this comparison, the $(3/2_3^-)$ state can be proposed as the confirmed state at 368.9(3) keV energy level.

 %, and also we are getting high $\log ft$ values for the allowed and unique transitions from the shell-model in the present case.}
 
  \begin{table}
		\centering
		\caption{Comparison between shell-model (SM) and experimental \cite{ zhu2008, singh2007, kondev2005,kondev2007, kondev2023, chiara2010, kondev2020, kondev2008, kondev2011, martin2007, chen2015,  morales2014} half-life values. Here, double brackets represent the unconfirmed states.}
  \label{half_life}
		\begin{ruledtabular}
			\begin{tabular}{lccccccccccc}
				{~~~~~Nucleus}& \multicolumn{2}{c}{Half-life}  \T\B \\
				\cline{2-3}

				&  Expt.  &SM \T\B\\\hline\T\B

				%$^{202}$Os($0^+$)& NA & 6.24 s\\
                    $^{202}$Os($0^+$)& NA & 5.50 s\\
			
				$^{202}$Ir($(1^-)$)&11(3) s \cite{zhu2008} & 582 s\\
				$^{202}$Ir($(2^-)$)&11(3) s \cite{zhu2008} & 5.32 $\times 10^3$ s\\
				$^{203}$Ir($(3/2^+)$)& NA & 9.29 s \\
			
				$^{199}$Pt($5/2^-$)&30.8(4) min \cite{singh2007}& 20.0 min\\
				$^{202}$Pt($0^+$)&44(15) h \cite{zhu2008} & 0.0574 h\\
				$^{204}$Pt($0^+$)&16$^{+6}_{-5}$ s \cite{morales2014} & 57.1 s \\
				%$^{196}$Au($2^-$)&6.1669(6) d & 406 d\\
				%$^{198}$Au($2^-$)&2.6941(2) d & 0.57 d\\
				$^{199}$Au($3/2^+$)&3.139(7) d \cite{singh2007}& 3.63 d\\
				$^{200}$Au($(1^-)$)  &48.4(3) min \cite{kondev2007} & 13.9 min\\
				$^{200}$Au$^{\rm m}$($12^-$)&18.7(5) h \cite{kondev2007} & 13.9 h\\
				$^{201}$Au($3/2^+$) &26.0(8) min \cite{kondev2023}& 20.1 min\\
				$^{202}$Au($(1^-)$)&28.4(12) s \cite{zhu2008} & 91.8 s\\
				$^{203}$Au($3/2^+$) &60(6) s \cite{kondev2005} & 81.7 s\\
				$^{204}$Au($(2^-)$) &39.8(9) s \cite{chiara2010}& 63.5 s\\
				$^{205}$Au($(3/2^+)$) &32.0(14) s \cite{kondev2020}& 26.5 s\\
				$^{203}$Hg($5/2^-$)&46.594(12) d \cite{kondev2005} & 66.5 d\\
				$^{205}$Hg($1/2^-$)  & 5.14(9) min \cite{kondev2020}& 5.13 min\\
				$^{206}$Hg($0^+$) &8.32(7) min \cite{kondev2008}& 7.90 min\\
				$^{207}$Hg($9/2^+$) &2.9(2) min \cite{kondev2011}& 2.61 min\\
				$^{204}$Tl($2^-$) &3.783(12) yr \cite{chiara2010}& 3.52 yr\\
				$^{206}$Tl($0^-$) &4.202(11) min \cite{kondev2008} & 3.80 min\\
				$^{207}$Tl($1/2^+$) &4.77(3) min \cite{kondev2011}& 4.93 min\\
				$^{208}$Tl($5^+$) &3.053(4) min \cite{martin2007}& 2.77 min\\
				$^{209}$Tl($1/2^+$) &2.162(7) min \cite{chen2015}& 2.05 min\\
				
			\end{tabular}
		\end{ruledtabular}
	\end{table}

 \begin{table*} 
		\centering
		\label{table4}
		\caption{Comparison between shell-model (SM) and experimental  $\log ft$ \cite{kondev2008,kondev2011,wennemann1994,farrelly2010} and average shape factor values for case (i) ($N= 126$) using $q_w = 0.41$,  $q_{\xi^\prime{v}} = 1.265$,  $q_x = 0.40$,  $q_u = 0.43$,  $q_z = 0.66$ in first forbidden transitions and $q_A=0.70$ for GT transitions. Here, double brackets represent the unconfirmed states.}\label{table4}
		\begin{ruledtabular}
			\begin{tabular}{lccccccccccc}
				\multicolumn{2}{c}{~~~~~Transition} &Energy& \multicolumn{2}{c}{$\log ft$} &\multicolumn{2}{c}{$ {[\overline{C(w_{e})}]}^{1/2}$} \T\B \\
				\cline{1-2}
				\cline{4-5}
				\cline{6-7}
				
				Initial ($J_i^\pi$)  & Final ($J_f^\pi$)  &(keV) &Expt.  &SM  & Expt. &SM \T\B\\\hline\T\B 
				
				$^{202}$Os($0^+$)&$^{202}$Ir($(1_1^-)$)&0.0&	NA&7.664 & NA &4.51 \\  
				&	$^{202}$Ir($(2_1^-)$)	& 0.0&NA&8.777 &NA & 1.25	\\

				$^{203}$Ir($3/2^+$)&$^{203}$Pt($1/2_1^-$)&0.0&NA &8.015 & NA& 3.01\\  
				&$^{203}$Pt($5/2_1^-$)&367.0&NA & 6.416&NA &19.0 \\

				$^{204}$Pt($0^+$)&$^{204}$Au($(2_1^-)$)&0.0& NA & 9.344& NA &0.652  \\

				$^{205}$Au($(3/2^+)$) \cite{wennemann1994,farrelly2010}&$^{205}$Hg($1/2_1^-$)& 0.0& 5.79(9) & 7.421 & 39.0(40)&5.96 \\  
				&$^{205}$Hg($5/2_1^-$)&379.16(21)  &6.37(12)	 &6.055 & 20.0(28) &28.7 \\  
				&$^{205}$Hg($3/2_1^-$)& 467.45(24)& 6.43(11) & 7.093 & 18.7(24) & 8.70\\ 
				&$^{205}$Hg($(1/2_2^-)$)& 1280.61(21)&5.51(12)  &8.258  & 53.8(74)& 2.28\\  
				&$^{205}$Hg($(3/2_2^-)$)& 1280.61(21)& 5.51(12) &5.175  & 53.8(74)& 79.2\\ 
				&$^{205}$Hg($(3/2_1^+)$)& 1280.61(21)& 5.51(12) & 9.680 &0.139(19) &0.00115 \\
				&$^{205}$Hg($(5/2_2^-)$)& 1280.61(21)& 5.51(12) &5.290  &53.8(74) & 69.4\\ 
				&$^{205}$Hg($(1/2_3^-)$)& 1325.08(24)& 5.53(11) & 8.772 &52.6(67) & 1.26\\  
				&$^{205}$Hg($(3/2_3^-)$)& 1325.08(24)& 5.53(11) & 5.251 &52.6(67) & 72.5\\ 
				&$^{205}$Hg($(3/2_2^+)$)& 1325.08(24)& 5.53(11) & 8.780 & 0.136(17)& 0.00323\\
				&$^{205}$Hg($(5/2_3^-)$)& 1325.08(24)& 5.53(11) & 5.997 &52.6(67) &30.7 \\ 
				&$^{205}$Hg($7/2_1^-$)& 1346.1(5)& NA & 8.416 &NA &1.90 \\ 
				%&$^{205}$Hg($9/2_1^-$)&3rd FNU& 1395.0(6)&  NA& 12.545 &NA &2438.317 \\ 
				&$^{205}$Hg($(1/2_4^-)$)& 1447.2(4)& 6.29(15) & 6.228 & 21.9(38)&23.6 \\  
				&$^{205}$Hg($(3/2_4^-)$)& 1447.2(4)& 6.29(15) &6.559 & 21.9(38)& 16.1\\ 
				&$^{205}$Hg($(3/2_3^+)$)& 1447.2(4)& 6.29(15) & 7.628 & 0.0568(98)& 0.0122\\
				&$^{205}$Hg($(5/2_4^-)$)& 1447.2(4)& 6.29(15) &6.300  &21.9(38) & 21.7\\ 
				&$^{205}$Hg($(5/2_1^+)$)& 1447.2(4)& 6.29(15) & 7.349 & 0.0568(98)&0.0168 \\

				$^{206}$Hg($0^+$) \cite{kondev2008}&$^{206}$Tl($0_1^-$) &0.0 &5.41(6)& 5.360 &60.4(42) &64.0\\
				&$^{206}$Tl($1_1^-$) &304.896(6) &5.24(10)&5.408 & 73.5(85) &60.5\\
				&$^{206}$Tl($1_2^-$)  &649.42(4) &5.67(8)&6.128 & 44.8(41) &26.4\\	
				
				$^{207}$Tl($1/2^+$) \cite{kondev2011}&$^{207}$Pb($1/2_1^-$)&0.0 &5.108(6)&5.175 &85.5(6) &79.2\\
				&$^{207}$Pb($5/2_1^-$) &569.64(10) &$>$10.4& 11.181 &$<$0.193 &  0.0786\\
				&$^{207}$Pb($3/2_1^-$) &897.76(10) &6.157(22) &7.138 & 25.6(65) & 8.26\\

			\end{tabular}
		\end{ruledtabular}
	\end{table*}

 	\subsubsection{\bf{Hg$\rightarrow$Tl}}
		The $\log ft$ values for Hg$\rightarrow$Tl chain have been calculated, which includes $\beta^-$ decay of $^{203,205,207}$Hg isotopes of this chain. For this chain, the $\log ft$  and average shape factor values calculated using the shell model match pretty well with the corresponding experimental results. The shell-model calculations for $^{203,205}$Hg isotopes have been performed with full model space as mentioned in the text, and shell-model calculations for $^{207}$Hg have been performed with the particle-hole excitations because of the involvement of core breaking during beta decay in the mentioned transitions. 
 For $^{203}$Hg ($5/2^-$) $\rightarrow$ $^{203}$Tl($3/2^+_1$) transition, the shell-model $\log ft$ value (6.661) matches quite well with the experimental $\log ft$ value, i.e., 6.457(8). Further moving towards $^{205}$Hg isotope, the shell-model $\log ft$ values for $^{205}$Tl($1/2_1^+$) and  $^{205}$Tl($5/2_1^+$) are 5.299 and 8.012 which are very close to the experimental $\log ft$ values, i.e., 5.257(11) and 8.28(22), respectively. Further, in the case of $^{207}$Hg($9/2^+$) $\rightarrow$ $^{207}$Tl $\beta$ decay at energy 3493.6(5) keV, three spin-parity states $(5/2^-,7/2^+,7/2^-)$ are possible experimentally. Thus, shell-model $\log ft$ values for all these spin-parity states are calculated separately, and it is inferred from the results that shell-model $\log ft$ value for the  $(7/2_1^+)$ and $(5/2_3^-)$ spin-parity states are closer to the experimental $\log ft$ value as compared to the $(7/2_5^-)$ state. Thus, the $(7/2_5^-)$ spin-parity state can be discarded at 3493.6(5) keV energy level. Similarly, at 3569.7(4) keV energy, two spin-parity states $(11/2^+,11/2^-)$ are possible experimentally, and shell-model $\log ft$ value for $(11/2_2^-)$ spin-parity state is closer to the experimental $\log ft$ value as compared to the other one. Thus, $(11/2_2^-)$ spin-parity state can be suggested at 3569.7(4) keV energy level.

	\subsubsection{\bf{Tl$\rightarrow$Pb}}
	In Tl$\rightarrow$Pb chain, the $\log ft$ values corresponding to $\beta$-decay of $^{204,206,208,209}$Tl are computed where case (i) model space is used for $^{204,206}$Tl and particle-hole excitation is used for $^{208,209}$Tl isotopes. Our shell-model $\log ft$ results agree fairly well with the experimental data. In the case of $^{204}$Tl(2$^-$)$\rightarrow$ $^{204}$Pb($0_1^+$) transition at ground state energy, the shell-model $\log ft$ value, i.e., 9.791 is close to the experimental $\log ft$ value, i.e., 9.7839(15). Further, in the case of $^{206}$Tl(0$^-$)$\rightarrow$ $^{206}$Pb($0_1^+$), the shell-model $\log ft$ value is 5.187, which is in excellent agreement with the experimental $\log ft$ value, i.e., 5.1775(13). Moving forward, in $^{208}$Tl, the shell-model $\log ft$ values for all transitions reasonably agree with the experimental data. Similarly, for $^{209}$Tl $\rightarrow$ $^{209}$Pb, the experimental and shell-model $\log ft$ values are fairly close to each other except for $^{209}$Pb$((5/2_1^{-}))$ and $^{209}$Pb$((3/2_4^{-}))$ states. 
 		
		\begin{figure*}
  \label{hl_plot}
			\includegraphics[width=85mm]{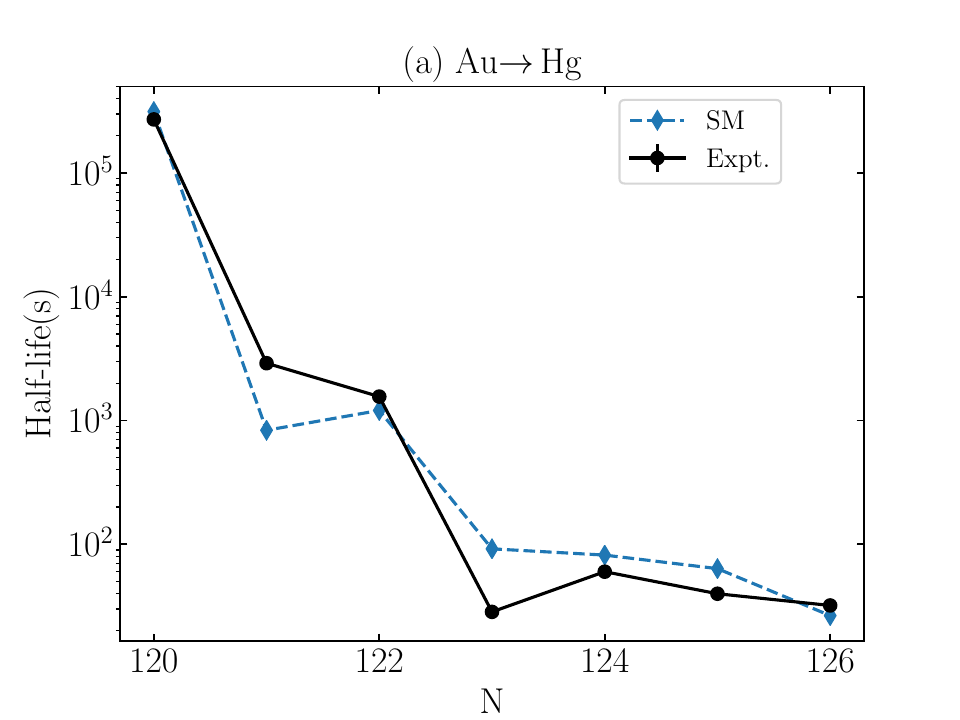}
			\includegraphics[width=85mm]{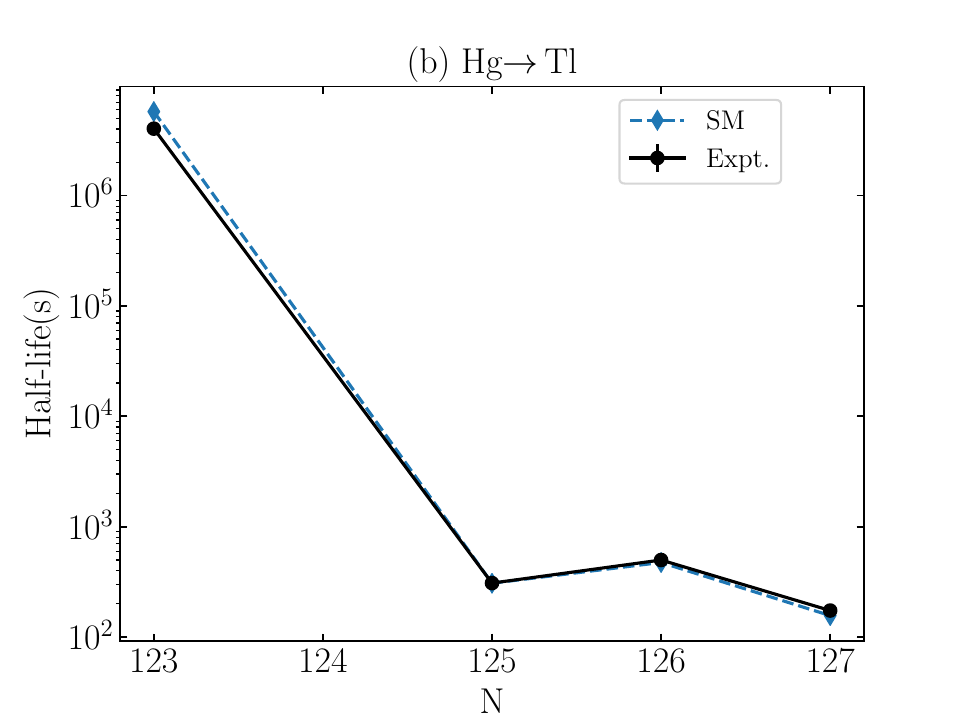}
			\includegraphics[width=85mm]{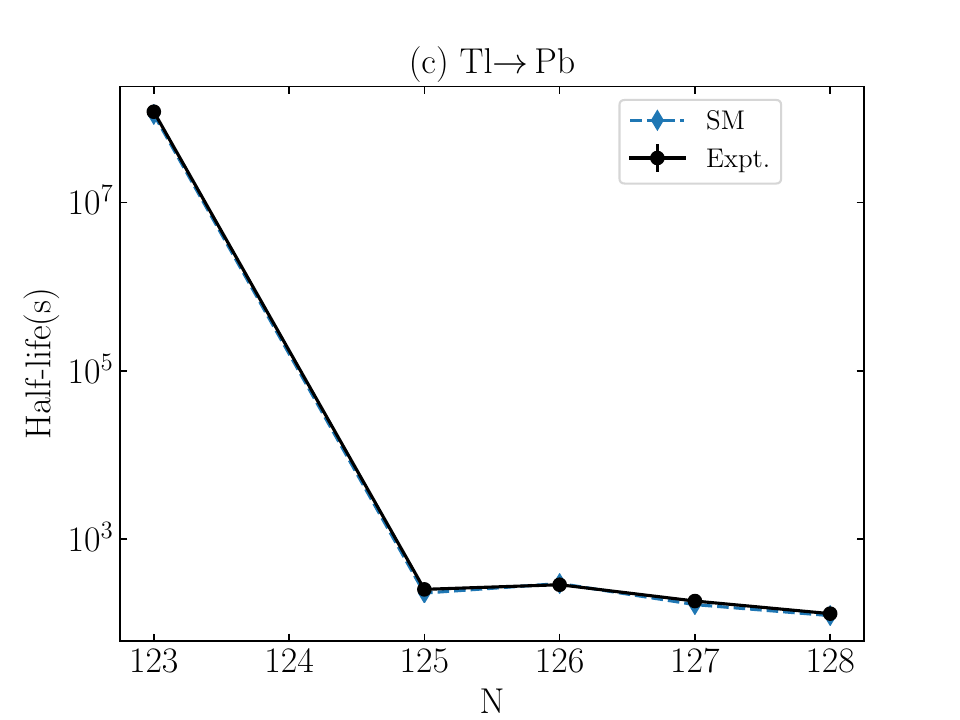}
			\caption{Comparison of calculated and experimental half-life values \cite{ zhu2008, singh2007, kondev2005,kondev2007, kondev2023, chiara2010, kondev2020, kondev2008, kondev2011, martin2007, chen2015} for  (a) Au, (b) Hg, (c) Tl isotopes. }\label{hl_plot}
		\end{figure*}

\begin{figure}\label{n126}
			\includegraphics[width=85mm]{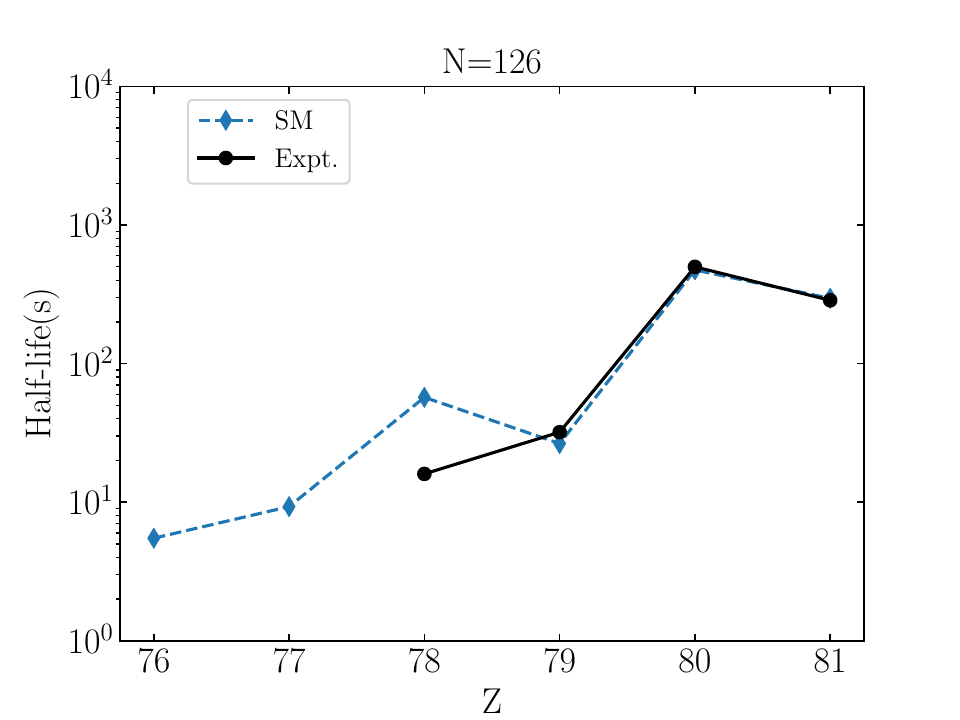}
			\caption{Comparison of calculated and experimental half-life values \cite{morales2014, kondev2008, kondev2011, kondev2020} for $N=126$ isotones. }\label{n126}
		\end{figure}

	\begin{figure*}
			\includegraphics[width=85mm]{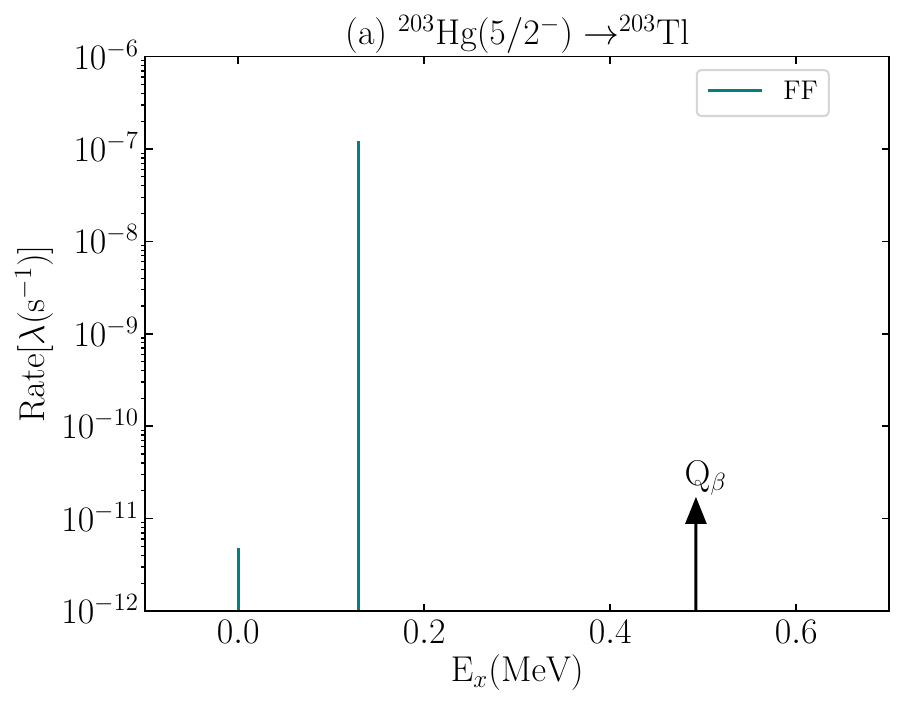}
			\includegraphics[width=85mm]{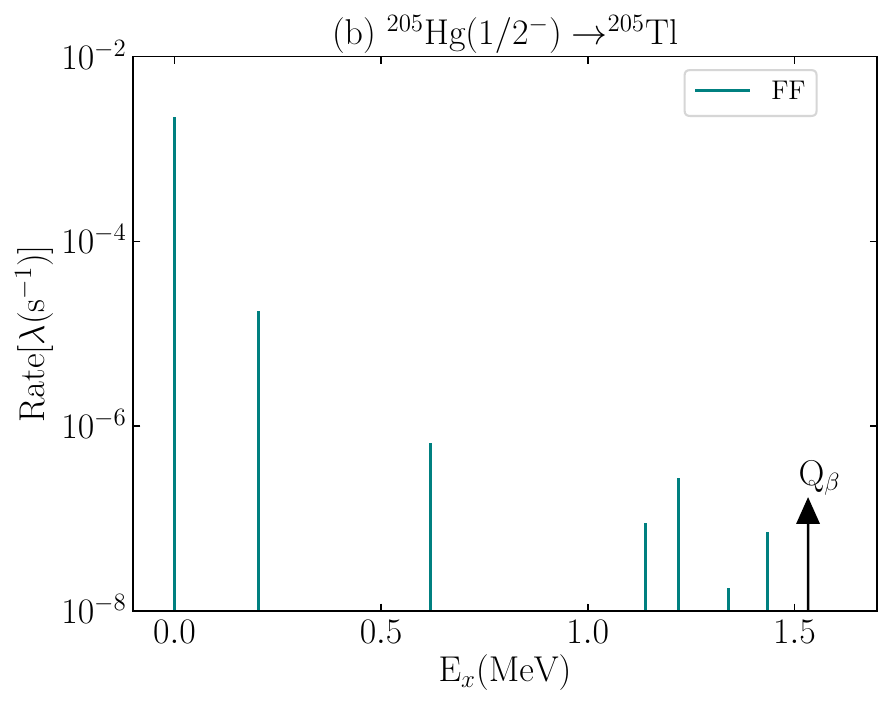}
			\includegraphics[width=85mm]{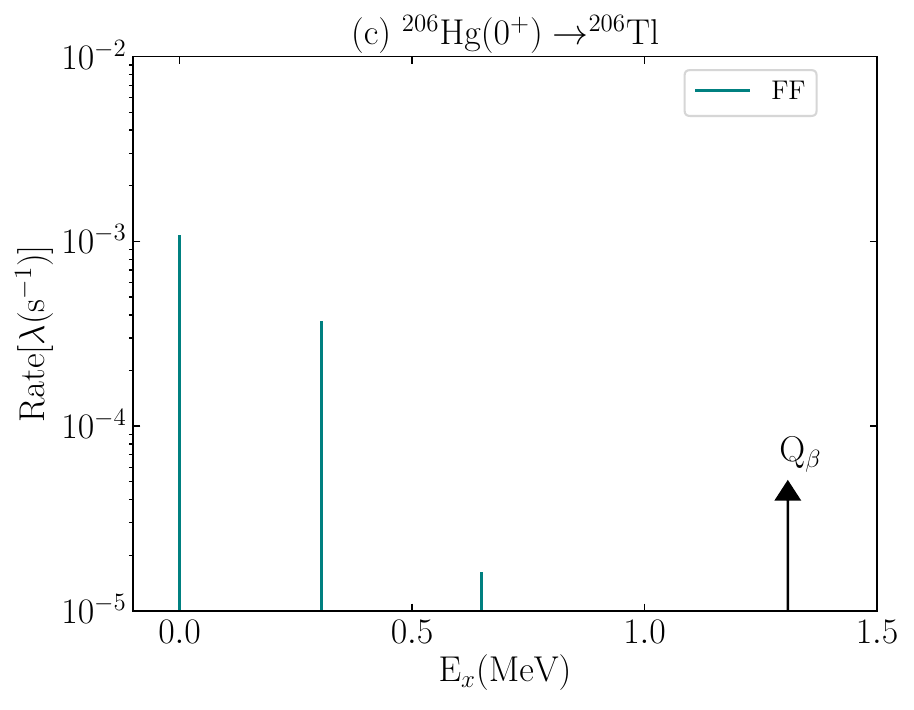}
			\includegraphics[width=85mm]{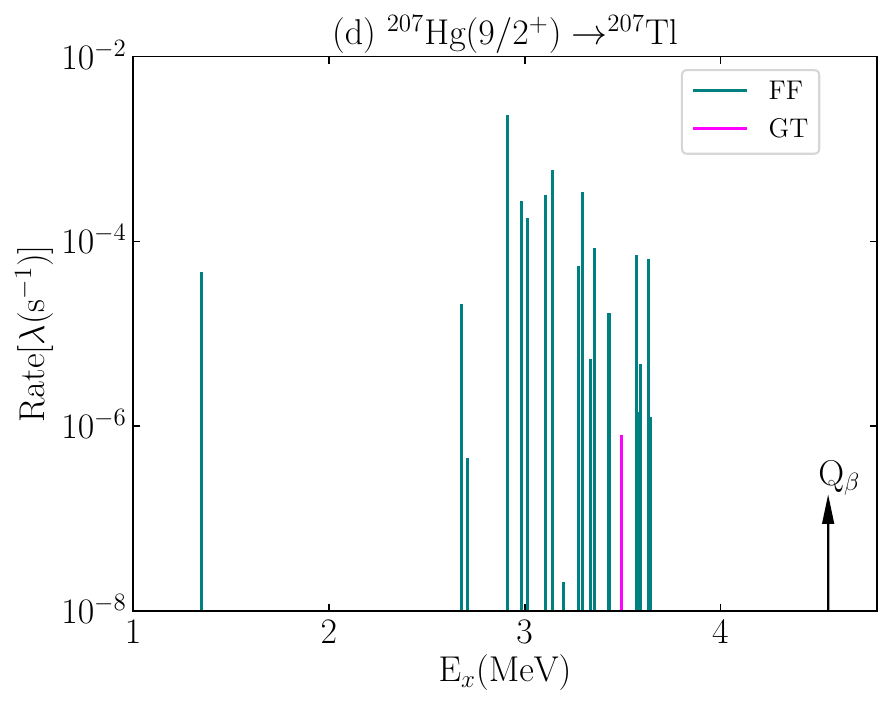}
			\caption{Decay rates [$\lambda$(s$^{-1}$)] vs. excitation energy [$E_x$ (MeV)] of daughter nucleus for Hg isotopes. }\label{fig:fig_3}
		\end{figure*}

		\begin{figure*}
			\includegraphics[width=85mm]{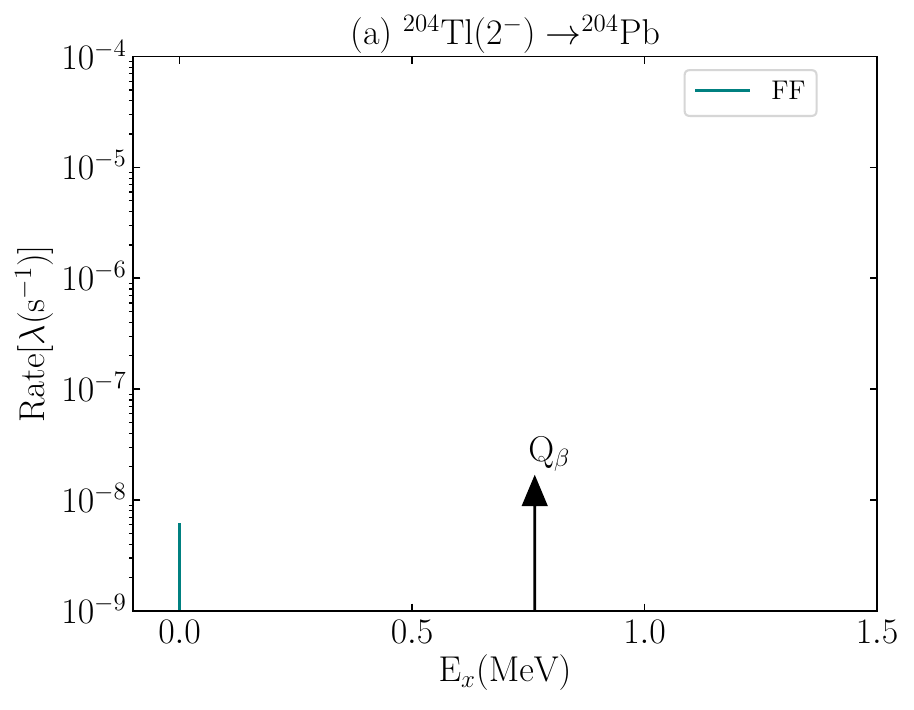}
			\includegraphics[width=85mm]{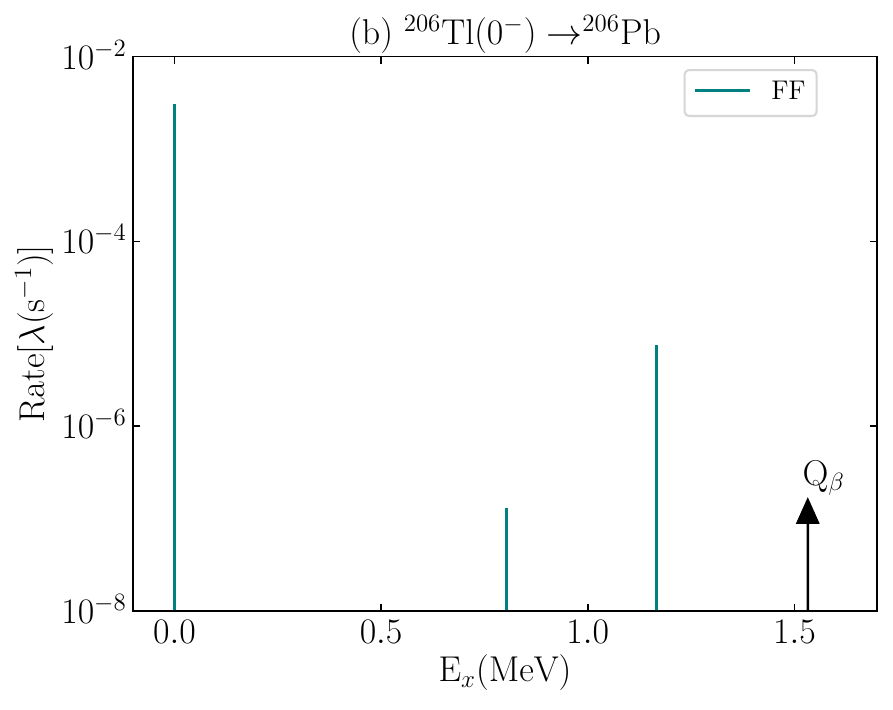}
			\includegraphics[width=85mm]{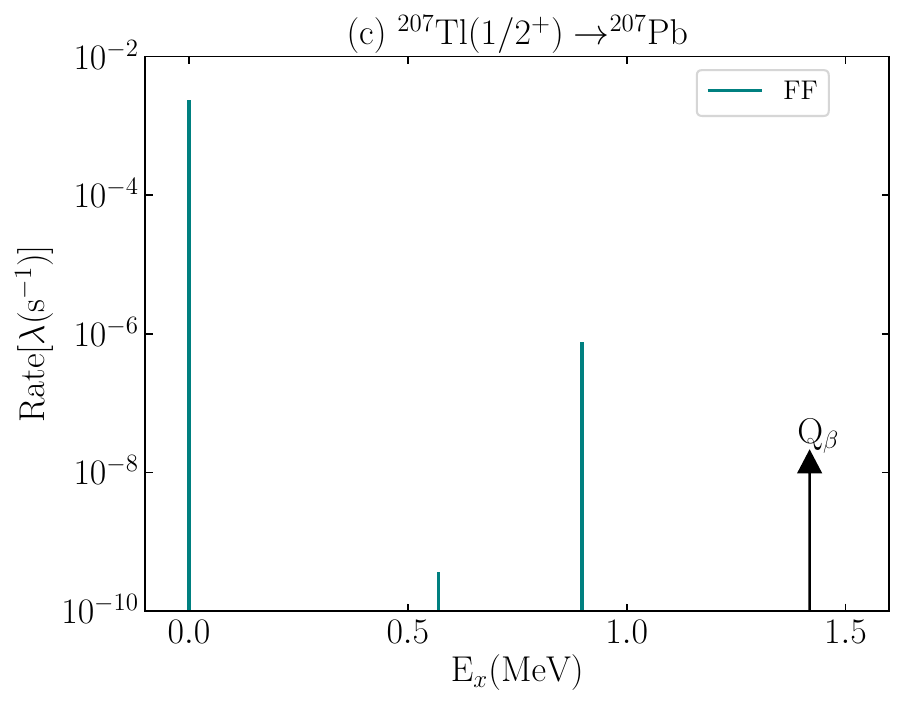}
			\includegraphics[width=85mm]{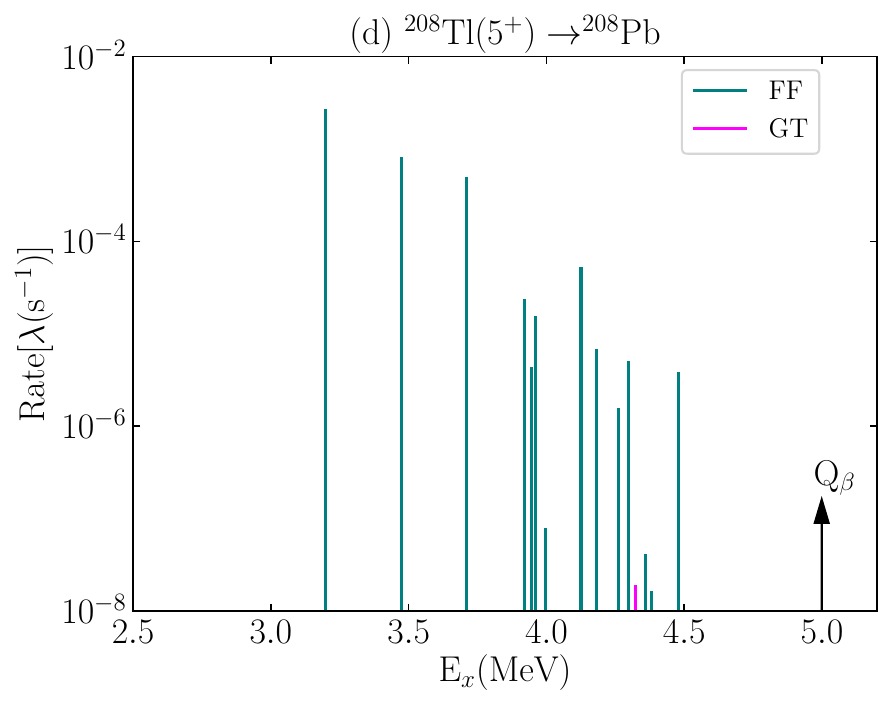}
			\includegraphics[width=85mm]{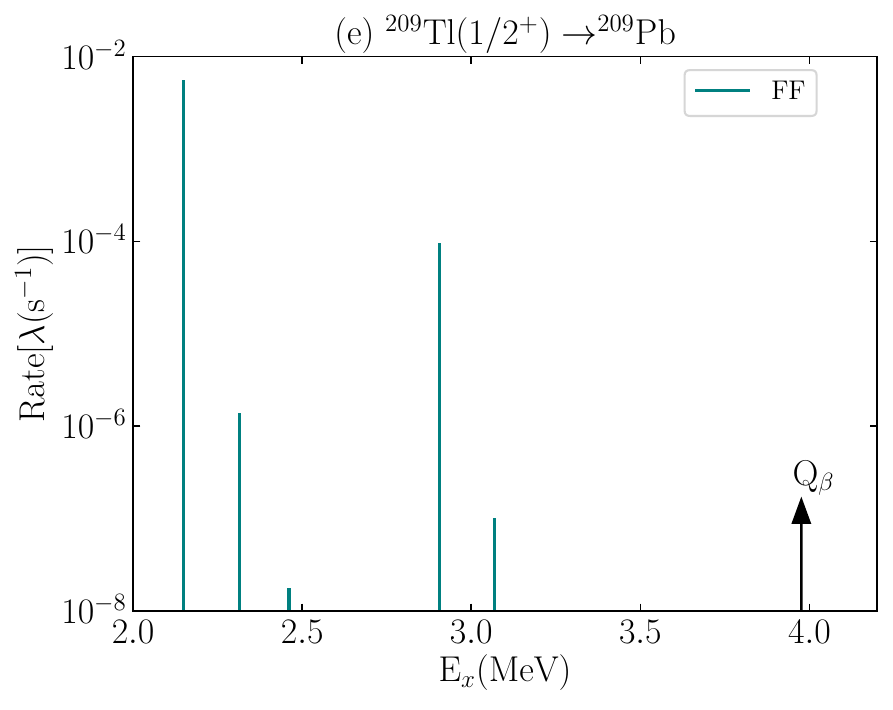}
			\caption{Decay rates [$\lambda$(s$^{-1}$)] vs. excitation energy [$E_x$ (MeV)] of daughter nucleus for Tl isotopes. }\label{fig:fig_4}
		\end{figure*}

\subsection{Half-life results}	
The half-life values for the various isotopic chains under consideration are listed in Table \ref{half_life}. Due to the extensive range of these half-life values, they are represented in  Fig. \ref{hl_plot} using a logarithmic scale of half-life in seconds, while they do not show the half-lives of the Os, Ir, and Pt isotopes and isomeric states. These plots show the neutron number for a particular isotopic chain on the $x$ axis and the half-life in the $\log$ scale on the $y$ axis. In the case of $^{202}$Os $\rightarrow$ $^{202}$Ir($J_f$) transitions, the ground state spin-parity of the daughter nucleus is not confirmed, and ($1^-$) and ($2^-$) states are tentatively assigned experimentally. Thus, the half-life value corresponding to both spin-parity states is calculated using shell-model calculations. However, the half-life value calculated by including ($1^-$) state is shown in the table, which comes out to be  5.50 s. Because, in $^{202}$Ir$\rightarrow$ $^{202}$Pt $\beta$ decay, the half-life for the $(1^-)$ state of  $^{202}$Ir is closer to the experimental one as compared to the other. Thus, $(1^-)$ state can be assigned as the g.s. of $^{202}$Ir. Further moving towards Pt$\rightarrow$Au isotopic chain, the shell-model and experimental half-life values are close to one another at $A=199,204$. Still, our shell-model calculations give a much lower half-life value at $A=202$ than the experimental one. Moving towards the Au$\rightarrow$Hg isotopic chain, our shell-model calculations for half-life values match pretty well with the experimental data. Similarly, for Hg$\rightarrow$Tl isotopic chain, the half-life values calculated using the shell model here also agree very closely with the corresponding experimental half-life values. In a similar manner, in the case of Tl$\rightarrow$Pb chain, the shell-model results are very good in accordance with the experimental data. This discussion leads to the conclusion that as one approaches the shell closure, shell-model half-life results get better.

\subsection{$N=126$ isotones} 
	Table \ref{table4} shows the $\log ft$ and average shape factor values for nuclei with neutron number $N=126$ for Os, Ir, Pt, Au, Hg, and Tl isotones. These shell-model results have been calculated via the inclusion of the quenching factors in the nuclear matrix elements. There are a few nuclei for which experimental data for $\beta^-$ decay results is unavailable. However, we have also calculated these nuclei's $\log ft$ values using shell-model calculations. In $^{202}$Os nucleus $\beta$ decay, two spin-parity states are tentatively proposed experimentally for the g. s. of the daughter nucleus. The shell-model $\log ft$ values for both spin-parity states have been computed. Further, the shell-model calculated $\log ft$ values agree pretty well with the experimental $\log ft$ values corresponding to $^{205}$Au, $^{206}$Hg and $^{207}$Tl isotones. For instance, in $^{206}$Hg$(0^+)\rightarrow$ $^{206}$Tl$(0^-_1)$ transition, the shell-model $\log ft$ value is 5.360 which is very close to the experimental $\log ft$ value, i.e., 5.41(6). Similarly, for $^{207}$Tl$(1/2^+)\rightarrow$ $^{207}$Pb$(1/2^-_1)$ transition, the shell-model $\log ft$ value is 5.175 which is close to the experimental $\log ft$ value, i.e., 5.108(6). Further, there are energy levels in $^{205}$Hg nucleus for which experimentally, more than one spin-parity state is assigned tentatively. The shell-model $\log ft$ values for all the tentatively assigned spin-parity states are compared with the corresponding experimental $\log ft$ value. For instance, four spin-parity states are tentatively proposed at the 1280.61(21) keV energy level. Based on the comparison with the experimental data, the $(1/2_2^-)$ and $(3/2_1^+)$ states can be discarded at 1280.61(21) keV energy level. Similarly,  $(3/2^-_3)$ can be proposed as the confirmed state at 1325.08(24) keV energy level, and $(3/2_3^+)$ can be discarded at  1447.2(4) keV energy level as $\log ft$ values for this state differ largely as compared to the other ones which are consistent with the experimental data. 
 
 The $\log ft$ values for $N=126$ isotones are also computed in Refs. \cite{suzuki2012,zhi2013}. In Ref. \cite{suzuki2012}, the $\log ft$ values using different sets of quenching factors in $g_A$ and $g_V$  are shown for $0^-$ and $1^-$ transitions corresponding to $Z=79,80$ and similarly, in Ref. \cite{zhi2013}, the $\log ft$ values are calculated corresponding to $Z=79,80,81$ nuclei using computed quenching factor values in first forbidden operators. Our $\log ft$ results are in reasonable agreement with these $\log ft$ values. The difference in the quenching factors between Ref. \cite{zhi2013} ($q_w = 0.66$,  $q_{\xi^\prime{v}} = 1.266$,  $q_x = 0.51$,  $q_u = 0.38$,  $q_z = 0.42$) and our work [case (i) values; see caption of Table \ref{table4} also] is due to the fact that we have taken huge set of data for the calculation of quenching factors in the first forbidden operators. Further, corresponding half-life values are reported in Table \ref{half_life}.
 %After calculating $\log ft$ values, 
 %the half-life values are calculated, which are included in Table \ref{half_life}. 
 In the half-life calculation of $^{202}$Os, $^{203}$Ir, and $^{204}$Pt $\beta$-decay, all possible allowed and FF transitions (at most five eigenvalues for each spin-parity) below the $Q$ value have been included for daughter nuclei using shell-model calculations, since there is not enough experimental data available for these transitions. These half-life values in seconds are plotted in Fig. \ref{n126}. As we can see from the plot, our shell-model half-life values match well with the available experimental data. The deviation can be seen at $Z=78$ because of the lack of enough experimental data for the $\beta$ decay of this isotone. 
	
 \subsection{Decay rates}
	
	Figures \ref{fig:fig_3} and \ref{fig:fig_4} show plots of partial decay rates, i.e., $\lambda(s^{-1})$ and the corresponding excitation energy [$E_x$(MeV)] of daughter nucleus for Hg and Tl isotopes. %These strength functions are calculated in order to evaluate the total half-life of the corresponding transition.
 The $Q$ value for corresponding $\beta$ decay is also shown in the plots.  These plots show the contribution of GT and first forbidden transitions in the total half-life of the nucleus. The GT strengths are mainly due to $\nu 0h_{9/2} \rightarrow \pi 0h_{11/2}$ transitions. In $^{208}$Pb region, the first forbidden transitions strongly compete with the GT transitions. In these plots, one can clearly see that GT transitions are generally observed at low transition energies. The contribution of GT transitions gets suppressed due to low transition energy and partial blocking (filling) of the $0h_{11/2}$ proton orbital. 
	
	\section{Summary and Conclusions} \label{Conclusion}
In this work, systematic shell-model calculations have been performed to evaluate the $\beta$-decay properties of nuclei in the south region of the $^{208}$Pb nucleus with $76 \leq Z \leq 82$. These $\beta$-decay properties include $\log ft$, average shape factor values, half-lives, and partial decay rates. 
Case (i) shell-model calculations have been performed corresponding to nuclei with $N\leq126$  and case (ii), i.e., particle-hole excitation for nuclei with $N>126$. 
The shell-model calculations for $\log ft$ values exhibit satisfactory agreement with the corresponding experimental data. We have thus shown that the present shell-model Hamiltonian is quite successful in describing the structure and transitions of nuclei in the south region of $^{208}$Pb. The Hamiltonian will be useful for the study of the structure of nuclei toward the ``blank spot". There are some unconfirmed states in the nuclei at various energy levels for which tentative spin-parity is assigned experimentally. Thus, we have compared the shell-model and the experimental $\log ft$ values for these states and suggested spin-parity accordingly corresponding to that particular energy level. 
Moreover, the $\log ft$ results are also calculated where experimental data are unavailable. Further, half-life values are calculated and plotted in logarithmic scale against neutron numbers (and proton numbers for $N=126$). One can conclude from the half-life plots that shell-model results improve as one moves towards shell closures, $Z=82$ and $N=126$. Further, the variations of the decay rates $[\lambda(s^{-1})]$ for $\beta$-decay transitions against the excitation energy of the daughter nucleus are also shown in the case of Hg and Tl isotopes.  Present study will be very useful for upcoming experimental data on $\beta$ decay properties in this mass region \cite{cern1,RIKEN1}.

 \section*{Acknowledgement}
 This work is supported by a research grant from SERB (India), CRG/2022/005167. S. S. would like to thank CSIR-HRDG (India) for the financial support for her Ph.D. thesis work. 
T. S. acknowledges JSPS (Japan) for a grant JSPS KAKENHI, Nos. JP19K03855 and JP20K03988. 
We acknowledge the National Supercomputing Mission (NSM) for providing computing resources of ‘PARAM Ganga’ at the Indian Institute of Technology Roorkee.
N.S. and A.K. acknowledge the support of ``Program for promoting researches on the supercomputer Fugaku'', MEXT, Japan (JPMXP1020230411) and the MCRP program of the Center for Computational Sciences, University of Tsukuba (NUCLSM). 

%	\newpage

\end{document}